\documentclass[iop, pdftex]{emulateapj}

\usepackage{amsmath}
\usepackage{amsfonts}
\usepackage{amssymb}
\usepackage{graphicx}
\usepackage{epsfig}
\usepackage[font=small, labelsep=space]{caption}  
\usepackage[caption=false]{subfig}

\def\JHU{1}
\def\STSCI{2}
\def\Nice{3}
\begin{document}

\slugcomment{\em Accepted in the Astronomical Journal}

\title{Polynomial Apodizers for Centrally Obscured Vortex Coronagraphs}
\author{Kevin Fogarty\altaffilmark{\JHU}, Laurent Pueyo\altaffilmark{\STSCI}, Johan Mazoyer\altaffilmark{\JHU, \STSCI}, Mamadou N'Diaye\altaffilmark{\Nice, \STSCI}}

\altaffiltext{\JHU}{Department of Physics and Astronomy, Johns Hopkins University, Baltimore, USA}
\altaffiltext{\STSCI}{Space Telescope Science Institute, Baltimore, USA}
\altaffiltext{\Nice}{Universit\'e C\^ote d'Azur, Observatoire de la C\^ote d'Azur, CNRS, Laboratoire Lagrange, UMR 7293, CS 34229, 06304 Nice Cedex 4}

\begin{abstract}
Several coronagraph designs have been proposed over the last two decades to directly image exoplanets. Among these designs, vector vortex coronagraphs provide theoretically perfect starlight cancellation along with small inner working angles when deployed on telescopes with unobstructed pupils. However, current and planned space missions and ground-based extremely large telescopes present complex pupil geometries, including large central obscurations caused by secondary mirrors, that prevent vortex coronagraphs from rejecting on-axis sources entirely. Recent solutions combining the vortex phase mask with a ring-apodized pupil have been proposed to circumvent this issue, but provide a limited throughput for vortex charges $>2$. We present pupil plane apodizations for charge 2, 4, and 6 vector vortex coronagraphs that compensate for pupil geometries with circularly symmetric central obstructions caused by on-axis secondary mirrors. These apodizations are derived analytically and allow vortex coronagraphs to retain theoretically perfect nulling in the presence of obstructed pupils. For a charge 4 vortex, we design polynomial apodization functions assuming a greyscale apodizing filter that represent a substantial gain in throughput over the ring-apodized vortex coronagraph design, while for a charge 6 vortex, we design polynomial apodized vortex coronagraphs that have $\gtrsim70\%$ total energy throughput for the entire range of central obscuration sizes studied.  We propose methods for optimizing apodizations produced with either greyscale apodizing filters or shaped mirrors. We conclude by demonstrating how this design may be combined with apodizations numerically optimized for struts and primary mirror segment gaps to design terrestrial exoplanet imagers for complex pupils.
\end{abstract}


\section{Introduction}

\renewcommand*{\thefootnote}{\arabic{footnote}}
\setcounter{footnote}{0}

Upcoming and proposed space and large ground-based telescope designs offer the sensitivity and resolution necessary to begin probing sub-Jovian and, in the case of space missions, terrestrial exoplanets with high contrast direct imaging \citep{2010Kasper_EELT, 2011Hinkley_Palomar, 2015Bolcar_LUVOIR, 2015Dalcanton_FutureUVOIR}. Recently, protoplanetary discs and massive, young exoplanets have been directly imaged using coronagraphs to suppress starlight to the level where these orbiting bodies can be observed \citep{2008Marois_Direct, 2008Kalas_Direct, 2009Lagrange_Direct, 2010Marois_Direct, 2011Andrews_Direct, 2013Carson_Direct, 2013Kuzuhara_Direct, 2013Rameau_Direct, 2013Oppenheimer_IRSpectra, 2015Mancini_Direct, 2015Macintosh_Direct, 2015Wagner_SPHERE, 2015Pueyo_HR8799}. However, observing fainter objects, such as Earth-like exoplanets, with future observatories poses a challenge for coronagraph designs, which will need to be adapted to the obstructed pupil geometries of telescopes with on-axis secondary mirrors and segmented primary mirrors\citep{ACAD, 2011Soummer, 2014Guyon}.

Several strategies are being developed to design coronagraphs for telescopes with obstructed pupils. Using one approach, the original Lyot coronagraph design can be modified by altering the shape of the Lyot stop to block artifacts introduced by the pupil geometry \citep{Anand_2001, 2005Sivar_Lyot}. Other approaches involve a combination of modifications to the pupil and focal plane geometries of the coronagraph, both to improve coronagraphic performance for unobstructed pupils and to compensate for complicated pupils. For example, the apodized pupil Lyot coronagraph (APLC) modifies the Lyot coronagraph by introducing a pupil apodizing mask in order to improve starlight suppression for broadband light with an obstructed pupil \citep{2005Soummer, 2015NDiaye_APLC}. 

Furthermore, the occulting spot in the Lyot coronagraph can be replaced by more complicated focal plane masks that involve a combination of phase shifting and apodization, such as the hybrid Lyot coronagraph \citep{2002Kuchner_Masks, 2007Moody_HybridLyot, 2008Moody_HybridLyotDemo, 2012Trauger_ComplexLyot, 2016Trauger_HybridWFIRST}.

Phase Induced Amplitude Apodization (PIAA) of the pupil plane with fixed shaped mirrors, combined with a system of apodizers and inverse-optics, has also been widely explored as a method for overcoming arbitrary pupil shapes \citep{2003Guyon, 2003Traub, 2005Guyon_PIAA, 2012Cady}. Fixed mirror apodization has also been combined with some of the complex coronagraph designs described above, such as the PIAA complex mask coronagraph (PIAACMC), which combines PIAA with a complex focal plane mask \citep{2014Guyon}.  These recent instrument designs depart considerably from the coronagraph proposed in \cite{Lyot_1939}, and represent substantial progress towards achieving terrestrial exoplanet direct imaging capability.

Coronagraphs using a vector vortex phase mask in the focal plane potentially provide substantial gains over other designs in terms of starlight suppression while maintaining high instrumental throughput for sources at close separations from a target star \citep{2005Foo, 2005Mawet}. In principle, the vector vortex coronagraph provides full cancellation of on-axis starlight in the absence of pupil obstructions with little suppression of nearby off-axis sources \citep{2011Mawet}. Vector vortex coronagraphs also permit small inner working angles (IWAs) (in some cases as small as $\sim 1 \lambda/D$), enabling observations of planets with relatively tight orbits \citep{2009Mawet, 2011Mawet_A}. In addition to entirely rejecting the light from a target star, these masks can be made theoretically achromatic by using subwavelength gratings to create phase shifts in the focal plane \citep{2005Mawet, 2007Mawet}. However, vector vortex coronagraphs only provide ideal starlight suppression with clear circular pupils, and are particularly impacted by the central obscuration imposed by a secondary mirror, which is the largest cause of pupil-geometry induced stellar flux residual in the coronagraphic image. \citep{2011Mawet, 2013Mawet_RAVC, 2014Fogarty_SAVC}.

Recent work has made progress adapting the vector vortex coronagraph to on-axis telescopes. \cite{2013Mawet_RAVC} proposed an analytical solution to the problem of the central obscuration by apodizing the pupil of the vector vortex coronagraph with a filter consisting of semi-transmissive, hard edged annuli or `rings'. The ring apodized vortex coronagraph (RAVC) solves the issue of suppression loss due to the central obscuration. Unfortunately, the throughput of the RAVC decreases rapidly as the radius of the central obscuration increases, so we seek alternatives to the RAVC for high-contrast imaging applications with on-axis telescopes.

In this paper we present a new method to compensate for the central obscuration of an on-axis telescope with a vortex coronagraph. Our results are a generalization of the RAVC, obtained by extending the formalism of previous work presented in \cite{2014Fogarty_SAVC} to derive novel solutions. Our previous work used linear programming to find apodization functions composed of basis Bessel functions that resolve the issue of limited throughput in the RAVC while providing $10^{-10}$ starlight suppression in monochromatic light. We adopt the linear programming formalism used in that paper to optimize pupil apodizations composed of piecewise polynomials. Using this new basis, we are able to use this approach to find throughput-maximizing pupil plane apodizations for vector vortex coronagraphs that produce theoretically perfect cancellation of on-axis starlight. 

Taking advantage of the the large number of possible solutions available that null the Lyot plane electric field, we can create apodizations for the polynomial apodized vortex coronagraph (PAVC) that are either discontinuous or smooth. We present PAVCs with topological charges 2, 4, and 6 that are optimized to maximize transmission if the pupil apodization is produced by an apodizing filter. We also present examples of apodization functions that may be produced by pairs of shaped mirrors, using a similar technique to PIAA, and that minimize the curvature of these mirrors. 

The PAVC designs we describe in this paper have several desirable properties that make them a good candidate for delivering extremely high contrast imaging on large, on-axis telescopes. Like the RAVC, PAVC designs are inherently broadband and offer an exact solution to the problem posed by centrally obscured pupils. However, since the RAVC `ring' apodization functions are a special case of PAVC apodizations, we are able to significantly improve throughput performance. We find that the throughput of a PAVC increases as the topological charge (the topological charge of the vortex is described in $\S$ 2.1) of the vortex increases, and that throughput falls as a function of secondary mirror radius more gradually than the RAVC for charges $> 2$. As a result, the PAVC is a viable option for designing a high-throughput vortex coronagraph on a telescope with a large central obscuration, particularly if the topological charge of the vector vortex is $>2$. By combining the PAVC with techniques to mitigate diffraction due to the discontinuities imposed on the pupil by secondary struts and primary mirror gaps, we can design intruments that deliver both high throughput and on-axis starlight suppression on telescopes with complex apertures.

\section{The Analytical Vortex Operator and Optimal Mask Algorithm}

\subsection{The Vector Vortex}

A vortex phase mask is an azimuthal phase ramp that replaces the occulting Lyot spot in the focal plane of a coronagraph. It causes the phase of an electric field propagating through it to undergo a rotation $e^{ic\theta}$, where $\theta$ is the angular coordinate in the coordinate system whose origin is the center of the phase mask and $c$ is the topological charge of the phase mask. The topological charge (hereafter referred to as `charge') determines the number of `rotations' the phase of the electric field undergoes per rotation around the center of the mask (see Figure 2 in \cite{2011Mawet_B}). A vortex phase mask in the focal plane of a coronagraph suppresses on-axis starlight by shifting it out of the radius of the original pupil in the subsequent Lyot plane.  Vortices always carry even-numbered charges, since odd-numbered charges do not suppress on-axis starlight in un-obscured pupils \citep{2005Mawet_Vortex}.

The charge of a vortex coronagraph affects both its IWA and how stable the instrument must be kept to prevent on-axis starlight from `leaking' into the final image. In particular, charge 2 vortex coronagraphs permit small, $\sim 1 \lambda/D$ IWAs, at the expense of sensitivity to low-order aberrations and finite stellar angular size \citep{2009Mawet, 2011Mawet_A}. Charge 2 vortex coronagraphs are suitable for instruments designed to achieve contrasts of $\sim 10^{-6}$; however, the sensitivity of the charge 2 vortex phase mask to low-order aberrations prevents it from being useful for obtaining larger contrasts \citep{2010Mawet_Aberrations}. Vortex coronagraphs with charges higher than 2, which offer lower sensitivity to low-order aberration at the expense of larger IWAs, are typically considered more likely candidates for terrestrial exoplanet characterizing instruments \citep{2008Jenkins, 2010Mawet_Aberrations}. 

The same property that allows a vortex phase mask in the focal plane to reject on-axis starlight from an unobscured pupil is what causes the vortex coronagraph to lose starlight suppression in the presence of a central obscuration \citep{2011Mawet, 2013Mawet_RAVC, 2014Fogarty_SAVC}. Namely, multiplying the Airy pattern of a uniform aperture in the focal plane by $e^{ic\theta}$ causes the light through that aperture to disperse into a ring with an inner radius corresponding to the outer radius of the aperture when it propagates into the Lyot plane. The amplitude of this ring peaks at the aperture radius, and falls off as the radius increases. However, a central obscuration in the pupil serves to subtract a uniform disc from the center of the pupil, essentially creating two concentric Airy patterns in the focal plane. Therefore, the vortex coronagraph disperses some of the on-axis starlight into a ring around the central obscuration in the Lyot plane, putting it back into the aperture of the pupil. Depending on the the size of the central obscuration, the residual starlight in the aperture limits the contrast achievable by the vortex coronagraph to $10^{-2}-10^{-1}$ if left uncorrected \citep{2011Mawet}.


\begin{table}[b!]
\footnotesize
\caption{Summary of Important Variables for Deriving the Apodized Vortex Formalism}
\label{table:Variables}
\vspace{5mm}
\centering
{
\begin{tabular}{lr}
\textit{Summary of Variables} \\
\hline
Description & Variable \\
\hline
\hline
Pupil/Lyot Plane Coordinates & $\left(r, \theta\right)$ \\
Focal Plane Coordinates & $\left(k, \theta\right)$ \\
Vortex Topological Charge & $c$ \\
Secondary Mirror Radius & $R_{S}$ \\
Primary Mirror Radius & $R_{P}$ \\
Inner Lyot Stop Radius & $R_{I}$ \\ 
Vortex Operator for a vortex of charge $c$ & $V_{c}$ \\
\end{tabular}  
\begin{flushleft}
\end{flushleft}  
}  
\end{table}

\subsection{Apodized Vortex Formalism}

\begin{figure*}
\begin{center}
\includegraphics{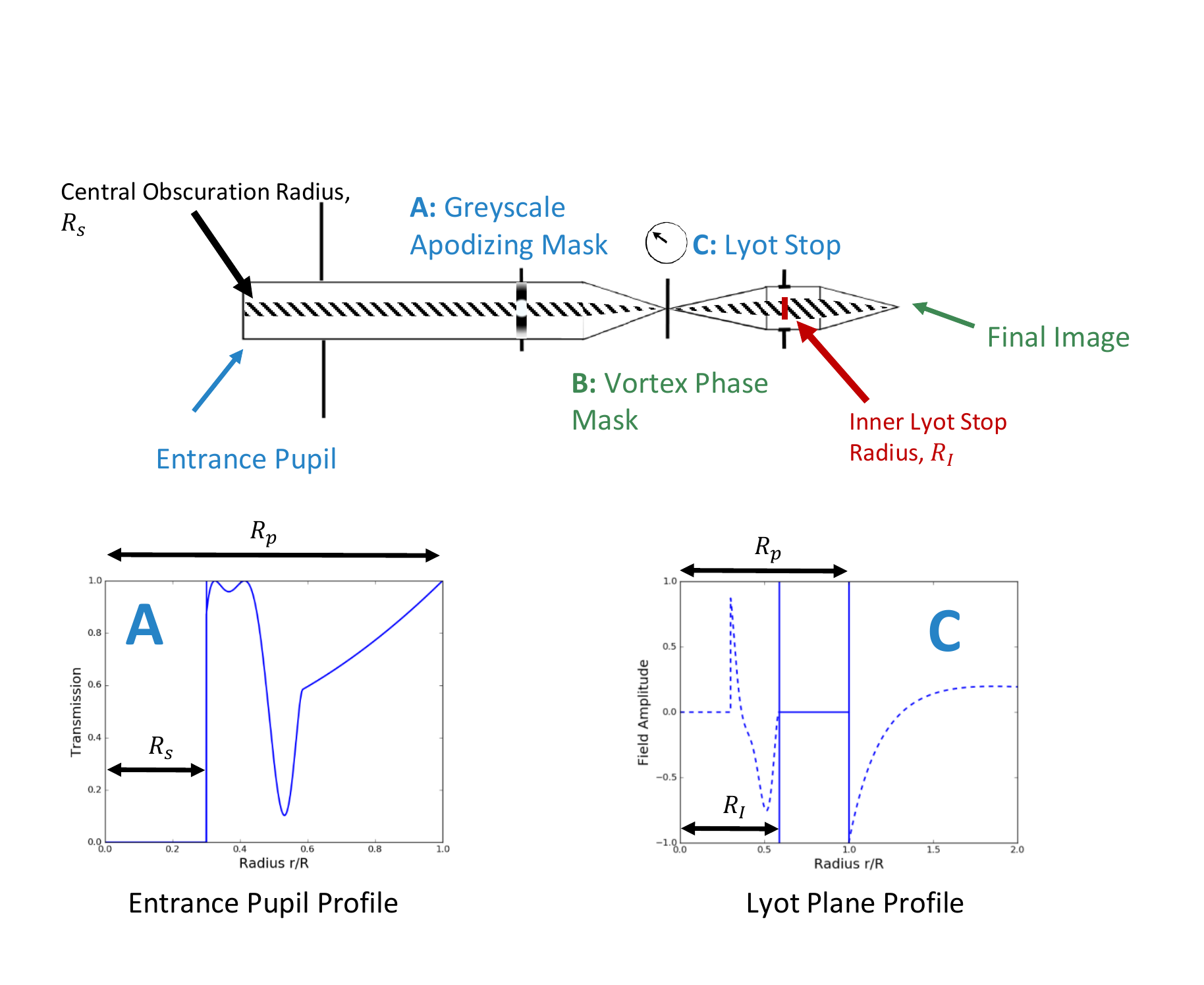}
\end{center}
\caption[]
{ \label{fig:Overview} An overview of the coronagraphic setup discussed in this paper. The coronagraph consists of three stages: (A) the apodizing mask, (B) the vortex phase mask, and (C) the inner Lyot stop. When the beam hits stages A and C, it is in the pupil plane, and at stage B it is in the focal plane. The diagonal hatching indicates the region of the beam that is blocked by the secondary mirror and by the inner Lyot stop. In this figure, blue text refers to parts of the beam in the pupil plane, while green text refers to parts of the beam in the focal plane, including the final image point spread function (PSF). The bottom left inset shows an example of the apodized transmission of flux from an on-axis source through stage A, with an apodization function optimized for a charge 4 vortex and a central obscuration that is 30$\%$ the radius of the pupil. The bottom right inset shows the field at stage C- the dashed lines show where the field is either outside the radius of the pupil $R_{P}$ or is blocked by the inner Lyot stop with radius $R_{I}$.}
\end{figure*}

We intend to solve the problem posed by telescope pupils with central obscurations by apodizing the entrance pupil of the vortex coronagraph. The coronagraphic setup we use is shown in Figure \ref{fig:Overview} and consists of three stages, which are labelled in the figure as Stages A, B, and C. Light from a directly on-axis source enters a pupil of radius $R_{P}$, which is obscured by a secondary mirror of radius $R_{S}$. At stage A, the pupil is apodized, which in Figure \ref{fig:Overview} is accomplished by using a greyscale apodizing filter. At stage B, the light passes through the vortex mask in the focal plane, before arriving the inner Lyot stop at stage C, which blocks flux within a radius $R_{I}$ in the Lyot plane. The final coronagraphic image is formed by focusing the beam in the Lyot plane onto a detector placed downstream of the coronagraph. Throughout this paper, we will use the polar coordinates $\left(r, \theta\right)$ when discussing the pupil and Lyot planes (stages A and C), and coordinates $\left(k, \theta\right)$ when discussing the focal plane (stage B) and final image (see Table \ref{table:Variables} for an overview of the important variables used in this discussion).

We define the ``vortex operator'' to be the operator mapping the electric field at the coronagraph entrance pupil to the electric field in the Lyot plane for the PAVC setup depicted in Figure \ref{fig:Overview}. We will show that when the pupil plane is apodized by a polynomial apodization function ($A\left(r\right) = c_{0} + c_{1}r + c_{2}r^{2} + ...$ at Stage A in Figure \ref{fig:Overview}), the vortex operator may be solved analytically. Using this analytical solution, we can constrain the coefficients of the polynomial apodization function to find optimal PAVCs that have no on-axis source flux in the Lyot plane using linear programming. 
 
We proceed to derive the vortex operator for a PAVC with a vortex of charge $c$ and a pupil apodization function $A\left(r\right)$, which we label $V_{c}\left[A\left(r\right)\right]$. Since the pupil radius in Figure \ref{fig:Overview} is $R_{P}$ and the radius of the central obscuration is $R_{S}$, the field in the pupil due to an on-axis point source of unit flux is 
\begin{equation}\label{eq:1}
P\left(r\right) = \Pi_{0, R_{P}}\left(r\right) - \Pi_{0, R_{S}}\left(r\right),
\end{equation}
where $\Pi_{a,b}\left(x\right)$ is the unit boxcar function with lower bound $a$ and upper bound $b$. Since the pupil is apodized by $A\left(r\right)$, the electric field at stage A, $F_{A}$, is
\begin{equation}\label{eq:2}
F_{A}\left(r\right) = A\left(r\right)P\left(r\right).
\end{equation}
At stage B, the beam undergoes a Fourier transformation, and is acted on by a vortex phase mask of charge $c$ in the focal plane. $F_{A}$ is an axisymmetric function, so the Fourier Transform of $F_{A}$ is equivalent to the zeroth order Hankel Transform, $H_{0}$. The field at stage B just before the vortex phase mask is therefore
\begin{equation}\label{eq:3}
F_{B}\left(k\right) = H_{0}\left[F_{A}\left(r\right)\right].
\end{equation}
After the beam passes through the vortex phase mask, the electric field is 
\begin{equation}
F_{B2}\left(k, \theta\right) = F_{B}\left(k\right)e^{ic\theta}.
\end{equation} 
The beam is transformed back to the pupil plane (referred to in this case as the Lyot plane) at stage C. Just before the Lyot stop, the field at stage C is $F_{C}\left(r\right) = H^{-1}_{c}\left[F_{B}\left(k\right)\right]$, where $H^{-1}_{c}$ is the inverse Hankel transform of order $c$. This can be obtained by re-arranging the inverse Fourier transform:
\begin{gather}\label{eq:4}
\begin{split}
F_{C}\left(r\right) &= \frac{1}{2\pi}\int{F_{B}\left(k\right)e^{ic\theta}e^{i\sin{kr\theta}}kdkd\theta_{k}} \\
&= \int_{0}^{\infty}{F_{B}\left(k\right)\left(\frac{1}{2\pi}\int_{-\pi}^{\pi}{e^{-i\left(c\tau - kr\sin\tau\right)} d\tau}\right)kdk}\\
&= \int_{0}^{\infty}{F_{B}\left(k\right)J_{c}\left(kr\right)kdk}\\
&= H_{c}^{-1}\left[F_{B}\left(k\right)\right],
\end{split}
\end{gather}
where $J_{c}$ is the Bessel function of order $c$ and where we have dropped the final phase term. Putting equations \ref{eq:2} through \ref{eq:4} together, we get the expression for the vortex operator: $V_{c}\left[A\left(r\right)\right] = H^{-1}_{c}\left[H_{0}\left[A\left(r\right)P\left(r\right)\right]\right]$. Expanding this out, we get,
\begin{equation}\label{eq:5}
\begin{split}
V_{c}\left[A\left(r\right)\right] &=  \int_{0}^{\infty}\left(\int_{0}^{\infty}A\left(r'\right)\left(\Pi_{0, R_{P}}\left(r'\right) -  \right. \right.\\ & \Pi_{0, R_{S}}\left(r'\right) \left. \right)J_{0}\left(kr'\right)r'dr' \left. \vphantom{\int} \right) J_{c}\left(kr\right)kdk \\
&= \int_{0}^{\infty}{\left(\int_{R_{S}}^{R_{P}}{A\left(r'\right)J_{0}\left(kr'\right)r'dr'}\right)J_{c}\left(kr\right)kdk}.
\end{split}
\end{equation}

\begin{figure*}
\includegraphics[height=11cm]{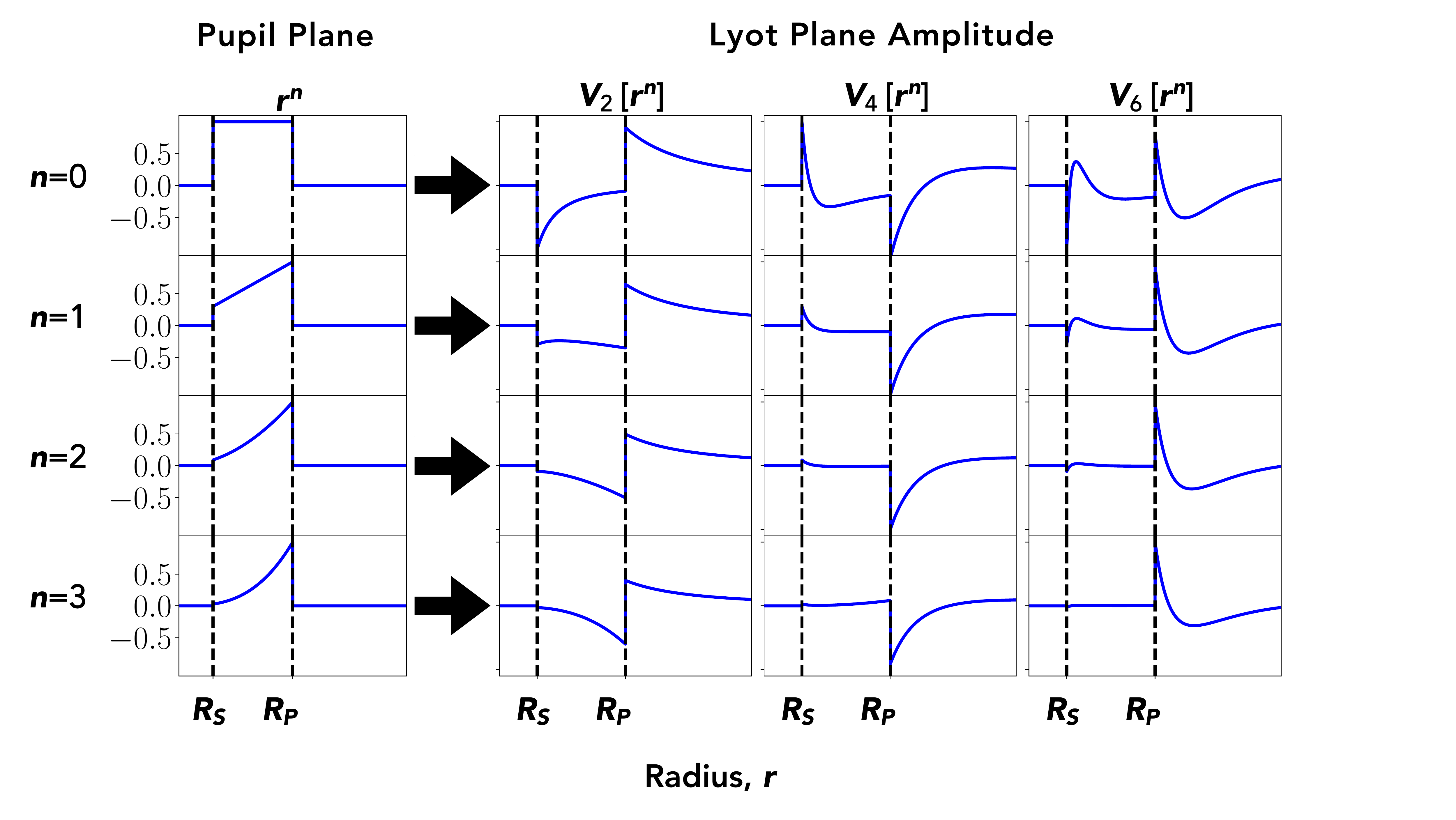}
\caption[]
{ \label{fig:Vc_rn} The vortex operator for a charge 2, 4, and 6 PAVC acting on $r^{n}$ for $n=0,1,2,3$. The left hand column shows the electric field amplitude in the pupil as a function of radius, apodized by (from top to bottom) $A\left(r\right) = 1$, $r$, $r^{2}$, or $r^{3}$. The plots in the 3x4 grid to the right depict the electric field amplitude in the Lyot plane as a function of radius after the pupil geometries in the left column are acted on by the vortex operator. The column depicts the the results of $V_{2}$ acting on $r^{n}$ for each value of $n$, the second column the results for $V_{4}$, and the third column for $V_{6}$. Solid blue lines depict electric field amplitude. The black dashed lines depict the positions of $R_{S}$ and $R_{P}$ in each plot.}
\end{figure*}

When the apodization function $A\left(r\right)$ is of the form $r^{n}$ (where $n$ is an integer), equation \ref{eq:5} takes on a simple form for even charges $c$. For $c=2$, we get,
\begin{equation}\label{eq:6}
V_{2}\left[r^{n}\right] = \begin{cases}
	0, & \text{if $r<R_{S}$} \\
	\frac{-n}{n+2}r^{n}-\frac{2}{n+2}R_{S}^{n+2}\frac{1}{r^{2}}, & \text{if $R_{S} \leq r \leq R_{P}$} \\
	\frac{-2}{n+2}\left(R_{S}^{n+2}-R_{P}^{n+2}\right)\frac{1}{r^{2}},& \text{if $r>R_{P}$}, \\
	\end{cases}
\end{equation}
for $c=4$,
\begin{equation}\label{eq:14}
V_{4}\left[r^{n}\right] = \begin{cases}
	0, & \text{if $r<R_{S}$} \\
	\frac{\left(n-2\right)nr^{n}}{\left(n+2\right)\left(n+4\right)} - \frac{4R_{S}^{n+2}}{n+2}\frac{1}{r^{2}}  & \text{if $R_{S} \leq r \leq R_{P}$} \\ + \frac{12R_{S}^{n+4}}{n+4}\frac{1}{r^{4}} \\
	-\frac{4\left(R_{S}^{n+2}-R_{P}^{n+2}\right)}{n+2}\frac{1}{r^{2}}+  & \text{if $r>R_{P}$} \\ \frac{12\left(R_{S}^{n+4}-R_{P}^{n+4}\right)}{n+4}\frac{1}{r^{4}} \\
	\end{cases}
\end{equation}
and for $c=6$,
\begin{equation}\label{eq:V6rn}
V_{6}\left[r^{n}\right] = \begin{cases}
0,  & \text{if $r<R_{S}$} \\
\frac{-n\left(n-2\right)\left(n-4\right)}{\left(n+2\right)\left(n+4\right)\left(n+6\right)}r^{n} - \frac{6R_{S}^{n+2}}{n+2}\frac{1}{r^{2}} & \text{if $R_{S} \leq r \leq R_{P}$} \\ + \frac{48R_{S}^{n+4}}{n+4}\frac{1}{r^{4}} - \frac{60R_{S}^{n+6}}{n+6}\frac{1}{r^{6}} \\
-\frac{6\left(R_{S}^{n+2}-R_{P}^{n+2}\right)}{n+2}\frac{1}{r^{2}}  & \text{if $r > R_{P}$} \\  + \frac{48\left(R_{S}^{n+4}-R_{P}^{n+4}\right)}{n+4}\frac{1}{r^{4}}  \\ - \frac{60\left(R_{S}^{n+6}-R_{P}^{n+6}\right)}{n+6}\frac{1}{r^{6}} \\
\end{cases}
\end{equation}
For larger even charges, similar expressions of increasing length may be calculated. We show $V_{c}\left[r^{n}\right]$ for $c=2,4,6$ and $n=0,1,2,3$ in Figure \ref{fig:Vc_rn}. Equations \ref{eq:6}-\ref{eq:V6rn} show that the vortex operator $V_{c}\left[A\left(r\right)\right]$ can be solved in closed form for any apodization function $A\left(r\right)$ that is in piecewise polynomial form (i.e. that can be decomposed into terms $r^{n}$ that are non-zero over some range of radii). We also note that analytical solutions to equation \ref{eq:5} for non-axisymmetric pupil geometries exist and are presented in Appendix A.

\subsection{Optimal Mask Algorithm}
By taking advantage of the simple expressions for $V_{c}\left[r^{n}\right]$ in Equations \ref{eq:6}-\ref{eq:V6rn}, we can set up a linear program to find optimal apodizing masks that null an on-axis point source observed with an obstructed pupil. We search for apodizations of the form:
\begin{equation}\label{eq:7}
A\left(r\right) = \begin{cases} \sum_{n=0}^{N}{a_{n}r^{n}}, &\text{if $R_{S} < r \leq R_{I}$}. \\
	\sum_{n=0}^{N}{a_{n}r^{n}+b_{n}r^{n}}, &\text{if $R_{I} < r \leq R_{P}$},
	\end{cases}
\end{equation}
where $R_{I} \geq R_{S}$ and where $N$ is the order of the piecewise polynomial we wish to use to describe $A\left(r\right)$.

Using the closed-form solutions we found for $V_{c}\left[r^{n}\right]$, we can find linear constraints on the coefficients $a_{n}$ and $b_{n}$ of the apodization function such that in the Lyot plane (Stage C in Figure \ref{fig:Overview}), the electric field is exactly zero at radii greater than $R_{I}$ and less than $R_{P}$. Since the inner Lyot stop at Stage C covers the region $r<R_{I}$, an apodization function that is a solution to this linear program produces perfect cancellation of the on-axis starlight.  

The apodization function in Equation \ref{eq:7} consists of two sets of components-- `$a_{n}$' components which are non-zero between $R_{S}$ and $R_{P}$ and zero elsewhere, and `$b_{n}$' components which are non-zero between $R_{I}$ and $R_{P}$ and zero elsewhere. The $b_{n}$ components can be thought of as seeing a central obscuration of radius $R_{I}$ instead of $R_{S}$. Therefore, when these components are acted on by the vortex operator in Equations \ref{eq:6}, \ref{eq:14}, or \ref{eq:V6rn}, $R_{S}$ is replaced by $R_{I}$. Defining $A\left(r\right)$ this way ensures that when we sum up the terms of $A\left(r\right)$ propagated into the Lyot plane, there exists a solution with zero electric field at $R_{I} < r \leq R_{P}$. 

For the case of a charge 2 vortex, the constraints that allow $V_{2}\left[A\left(r\right)\right] = 0$ in $R_{I} < r \leq R_{P}$ are:
\begin{subequations}\label{eq:C2Null}
\begin{equation}
\sum_{n=0}^{N}{\left(-\frac{2R_{S}^{n+2}}{n+2}a_{n}-\frac{2R_{I}^{n+2}}{n+2}b_{n}\right)} = 0,
\end{equation}
\begin{equation}
a_{n} + b_{n} = 0, \text{if $n > 0$}.
\end{equation}
\end{subequations}
For a charge 4 vortex, the constraints are 
\begin{subequations}\label{eq:C4Null}
\begin{equation}
  \sum_{n=0}^{N}{\left(- \frac{4R_{S}^{n+2}}{n+2}a_{n} - \frac{4R_{I}^{n+2}}{n+2}b_{n}\right)} = 0,
\end{equation}
\begin{equation}
  \sum_{n=0}^{N}{\left(\frac{12R_{S}^{n+4}}{n+4}a_{n} + \frac{12R_{I}^{n+4}}{n+4}b_{n}\right)} = 0,
\end{equation}
\begin{equation}
a_{n} + b_{n} = 0, \text{if $n \neq \left(0, 2\right) $}.
\end{equation}
\end{subequations}
For a charge 6 vortex, they are
\begin{subequations}\label{eq:C6Null}
\begin{equation}
\sum_{n=0}^{N}{\left(-\frac{6R_{S}^{n+2}}{n+2}a_{n} - \frac{6R_{I}^{n+2}}{n+2}b_{n}\right)} = 0,
\end{equation}
\begin{equation}
\sum_{n=0}^{N}{\left(\frac{48R_{S}^{n+4}}{n+4}a_{n} + \frac{48R_{I}^{n+4}}{n+4}b_{n}\right)} = 0,
\end{equation}
\begin{equation}
\sum_{n=0}^{N}{\left(-\frac{60R_{S}^{n+6}}{n+6}a_{n} - \frac{60R_{I}^{n+6}}{n+6}b_{n}\right)} = 0,
\end{equation}
\begin{equation}
a_{n} + b_{n} = 0, \text{if $n \neq \left(0, 2, 4\right) $}.
\end{equation}
\end{subequations}
It is also necessary that the solution correspond to a physically real apodizing mask, which is guaranteed  by the additional constraint
\begin{equation}\label{eq:12}
0 \leq A\left(r\right) \leq 1.
\end{equation}

If we wish to produce $A\left(r\right)$ using a classical apodizing mask, the constraints in either Equations \ref{eq:C2Null}, \ref{eq:C4Null}, or \ref{eq:C6Null}, and in Equation \ref{eq:12} are sufficient. However, if we wish to produce $A\left(r\right)$ using shaped mirrors, we also desire mask shapes that are smooth. We can constrain mask shapes to be continuous, using
\begin{equation}\label{eq:10}
\sum_{n=0}^{N}{b_{n}R_{I}^{n}} = 0,
\end{equation}
and smooth, using
\begin{equation}\label{eq:11}
\sum_{n=0}^{N}{nb_{n}R_{I}^{n-1}} = 0.
\end{equation}
It is interesting to note that since we are free to add degrees of freedom to the problem by increasing the order $N$ of the piecewise polynomial describing $A\left(r\right)$, we can continue to add linear constraints while ensuring that the problem remains solvable. For example, we can ensure that the mask solution be $C^{k}$ smooth for any order $k$.

Having specified the constraints on $A\left(r\right)$, it remains choose a figure-of-merit (FOM)  in order to calculate an optimal apodization function. The choice of FOM depends on both instrument performance goals and how the apodization function is produced. One of the key features of the PAVC is that it can be optimized according to different combinations of FOM and constraints.

 For a PAVC with $A\left(r\right)$ produced with a classical apodizing filter, the FOM ought to maximize filter transmission, while the smoothness and continuity constraints can be neglected. Meanwhile, for a PAVC with shaped mirrors, a FOM must be selected that minimizes mirror curvature. In the following sections, we motivate the FOMs for PAVCs produced with either apodizing filters or with shaped mirrors, and demonstrate the results we obtain for a range of central obscuration sizes.

\vspace{1.5mm}
\begin{center} \textasteriskcentered \textasteriskcentered \textasteriskcentered \end{center}
\vspace{1.5mm}
To summarize, the linear program that needs to be solved to find a perfectly nulling apodizing pupil mask is defined by:
\newline
\textbf{\textbullet \ \  A figure of merit (FOM) which depends on the desired properties of the coronagraph,} 
\newline
and by the set of constraints that ensure:
\newline
\textbf{\textbullet \ \ Lyot plane nulling (equations \ref{eq:C2Null}, \ref{eq:C4Null}, or \ref{eq:C6Null})}
\newline
\textbf{\textbullet \ \ Feasibility (equation \ref{eq:12})}
\newline
\textbf{\textbullet \ \ Continuity (equation \ref{eq:10}) (optionally)}
\newline
\textbf{\textbullet \ \ Smoothness (equation \ref{eq:11}) (optionally)}.
\newline

\section{The Apodizing Filter PAVC}

\subsection{Figure of Merit Selection}

The apodization function for a PAVC incorporating an apodizing mask/filter at Stage A in Figure \ref{fig:Overview} ought to be optimized to permit as much flux through the filter as possible. Therefore, the linear FOM we use is the transmission through the mask,
\begin{equation}\label{eq:13}
T = \int_{R_{I}}^{R_{P}}A\left(r\right)rdr.
\end{equation}
To calculate the FOM,  we only integrate $T$ between $R_{I}$ and $R_{P}$ since light inside $R_{I}$ is blocked by the inner Lyot stop. 

For a charge 2 PAVC, maximizing $T$ is equivalent to maximizing total energy throughput (defined as the integrated energy in the Lyot plane that is transmitted through the pupil apodizing filter and inner Lyot stop; see Table \ref{table:Throughput_Defs} for an overview of how `throughput' is defined and used throughout this paper), since given the constraints in Equation \ref{eq:C2Null}, only the $a_{0}$ and $b_{0}$ terms in $A\left(r\right)$ do not sum to zero in the region $R_{I}<r<R_{P}$. We emphasize that $T$ is not the same quantity as total energy throughput, which is a non-linear function of the apodization. $T$ in this case is
\begin{equation}
 T = \frac{a_{0}+b_{0}}{2}\left(R_{P}^{2}-R_{I}^{2}\right),
\end{equation}
while the total energy throughput is
\begin{equation}
 T.E. = \frac{\left(a_{0} + b_{0}\right)^{2}}{2}\left(R_{P}^{2}-R_{I}^{2}\right),
\end{equation}
so subject to the constraint that $a_{0} + b_{0} \geq 0$, it can clearly be seen that finding the solution to the linear program that maximizes $a_{0} + b_{0}$ will optimize both $T$ and total energy throughput.

This is not the case for higher charges, so strictly speaking optimizing $T$ will not allow us to find the maximum total energy throughput attainable with a classically apodized $c > 2$ PAVC. To find the maximum total energy throughput achievable by a charge 4 or higher PAVC for a given $R_{S}$, one would have to use a non-linear optimizer rather than the linear programming routine we use here. For this reason, maximizing transmission instead of throughput is a commonly used method to linearize problems involving optimizing apodized coronagraph designs \citep{2003Vanderbei_Masks, 2012Carlotti_Masks, 2015NDiaye_APLC}. Therefore, we use $T$ as the FOM for charge 4 and 6 PAVCs, while bearing in mind that we may overlook solutions to Equations \ref{eq:C4Null} and \ref{eq:C6Null} that produce higher total energy throughputs than we report. 

\begin{table*}[t!]
\footnotesize
\caption{Definitions of Throughput}
\label{table:Throughput_Defs}
\vspace{5mm}
\centering
{
\textit{Definitions of Throughput for Coronagraphic Instruments used in $\S$3-5.} \\
\begin{tabular}{p{35mm}|p{100mm}}
\hline
Name & Definition \\
\hline
\hline
\flushleft Total Energy Throughput & \begin{flushleft} The integrated energy in the Lyot plane of a PAVC that is transmitted through the pupil and Lyot stop. For a classically apodized PAVC, this quantity can be calculated by
\begin{equation}
T.E. = \frac{2\pi\int_{R_{I}}^{R_{P}}{A^{2}\left(r\right)rdr}}{\pi\left(R_{P}^{2} - R_{S}^{2}\right)}.
\end{equation} \end{flushleft} \\
\hline
\flushleft Total Energy Off-Axis Throughput & \begin{flushleft} The total energy of an off-axis source in the Lyot plane. For a classicaly apodized PAVC, this quantity can be calculated by
\begin{equation}
T.E.\left(r\right) = \frac{\int_{R_{I}}^{R_{P}}{\int_{0}^{2\pi}{\left(V_{c}\left[A\left(r\right)F\left(r,\theta\right)\right]\right)^{2}d\theta}rdr}}{\pi\left(R_{P}^{2}-R_{S}^{2}\right)},
\end{equation}
where $F\left(r, \theta\right)$ is the off-axis source flux in the telescope pupil. \end{flushleft} \\
\hline
\flushleft Encircled Energy Throuhput & \begin{flushleft} The integral of the energy contained in an aperture of diameter 0.7$\lambda$/D centered on the off-axis source position in the image plane of the PAVC. If $\left(K_{x}, K_{y}\right)$ is the position of the off-axis source in coordinates centered on the on-axis star, then the encircled energy throughput is
\begin{equation}
E.E.\left(K_{r}\right) = \frac{\int_{\left(k_{x}-K_{x}\right)^{2} + \left(k_{y}-K_{y}\right)^{2} \leq \left(0.35\lambda/D\right)^{2}}{I^{2}\left(k_{x}, k_{y}\right)dk_{x}dk_{y}}}{\pi\left(R_{P}^{2}-R_{S}^{2}\right)},
\end{equation}
where $K_{r} \equiv \sqrt{K_{x}^{2}+K_{y}^{2}}$ and $I\left(k_{x},k_{y}\right)$ is the amplitude in the image plane. This quantity approximates the relative signal strength that would be obtained for an off-axis source when performing aperture photometry on the source position in the image plane.\end{flushleft} \\
\hline
\end{tabular}  
\begin{flushleft}
\end{flushleft}  
}  
\end{table*}

\begin{figure*}
\centering
\subfloat{\includegraphics[height=6.5cm]{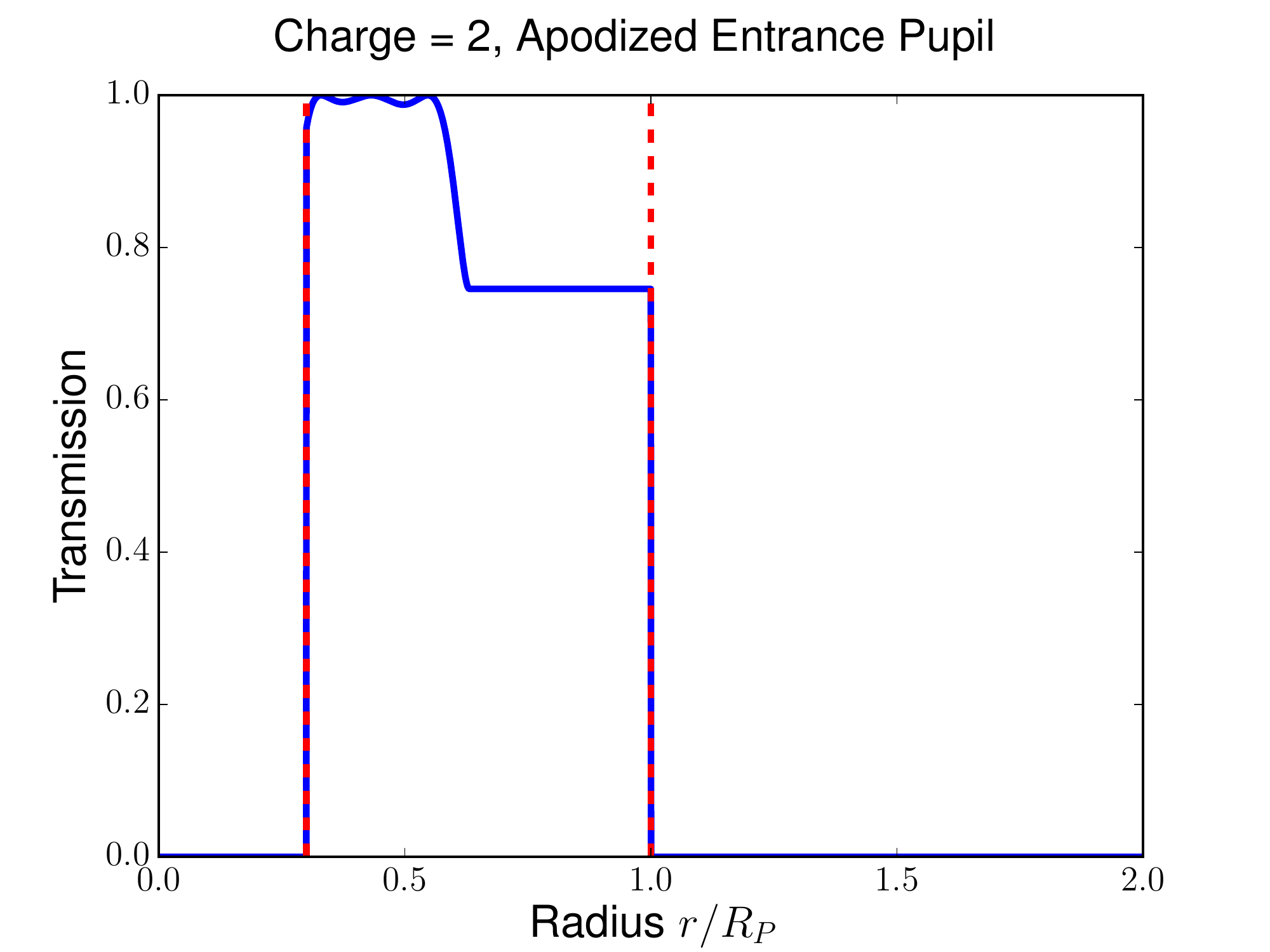}}
~
\subfloat{\includegraphics[height=6.5cm]{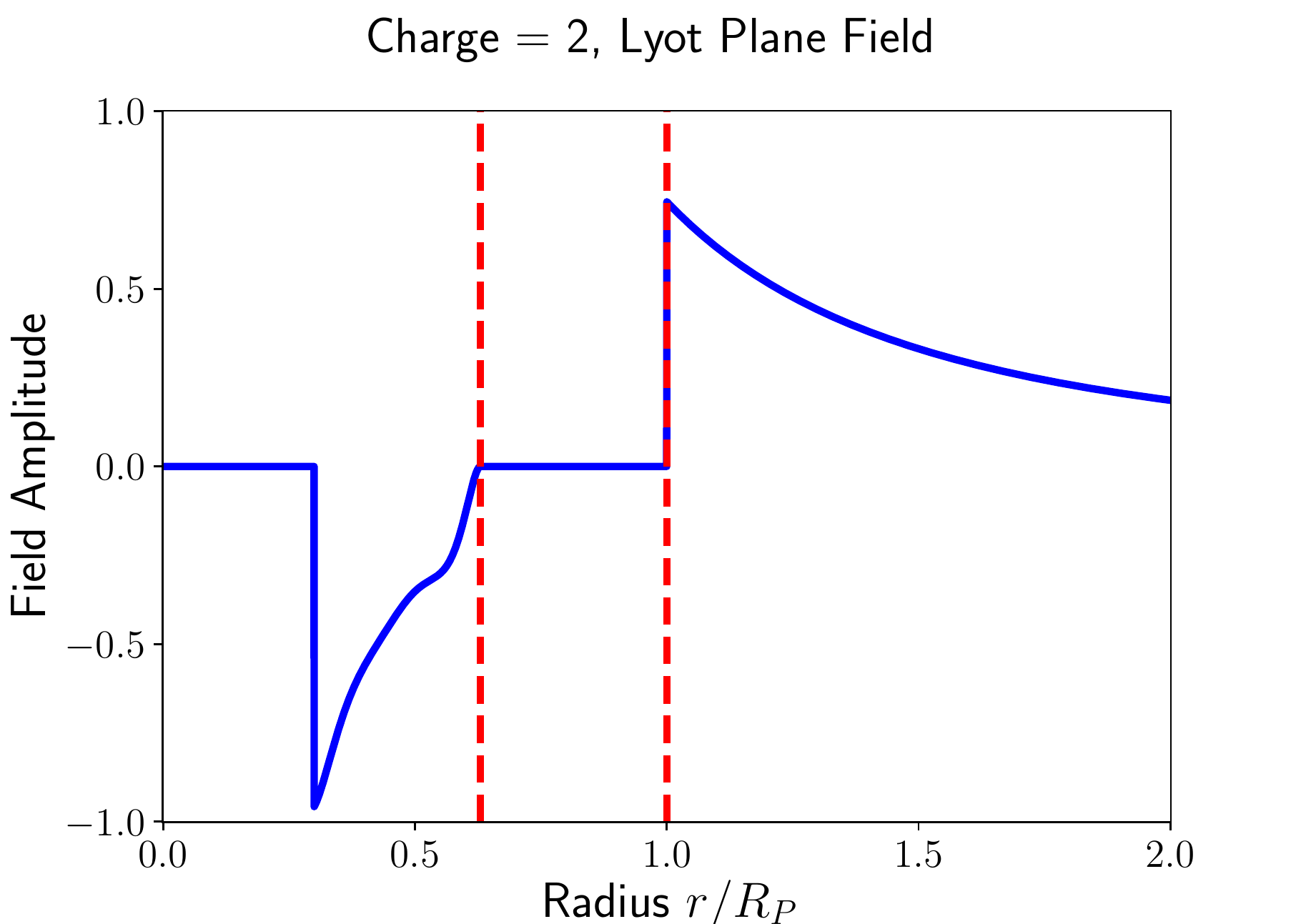}}
\newline
\subfloat{\includegraphics[height=6.5cm]{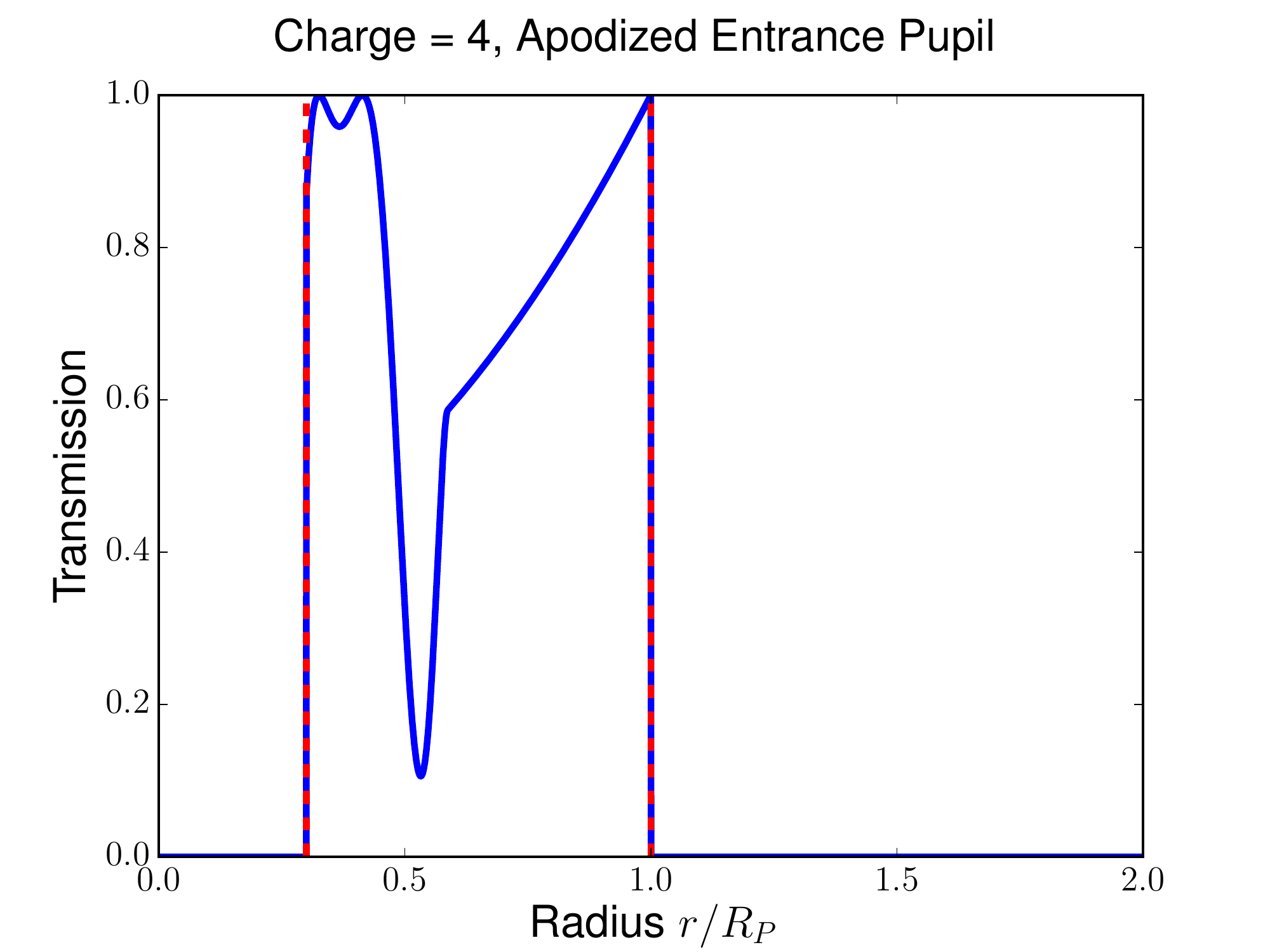}}
~
\subfloat{\includegraphics[height=6.5cm]{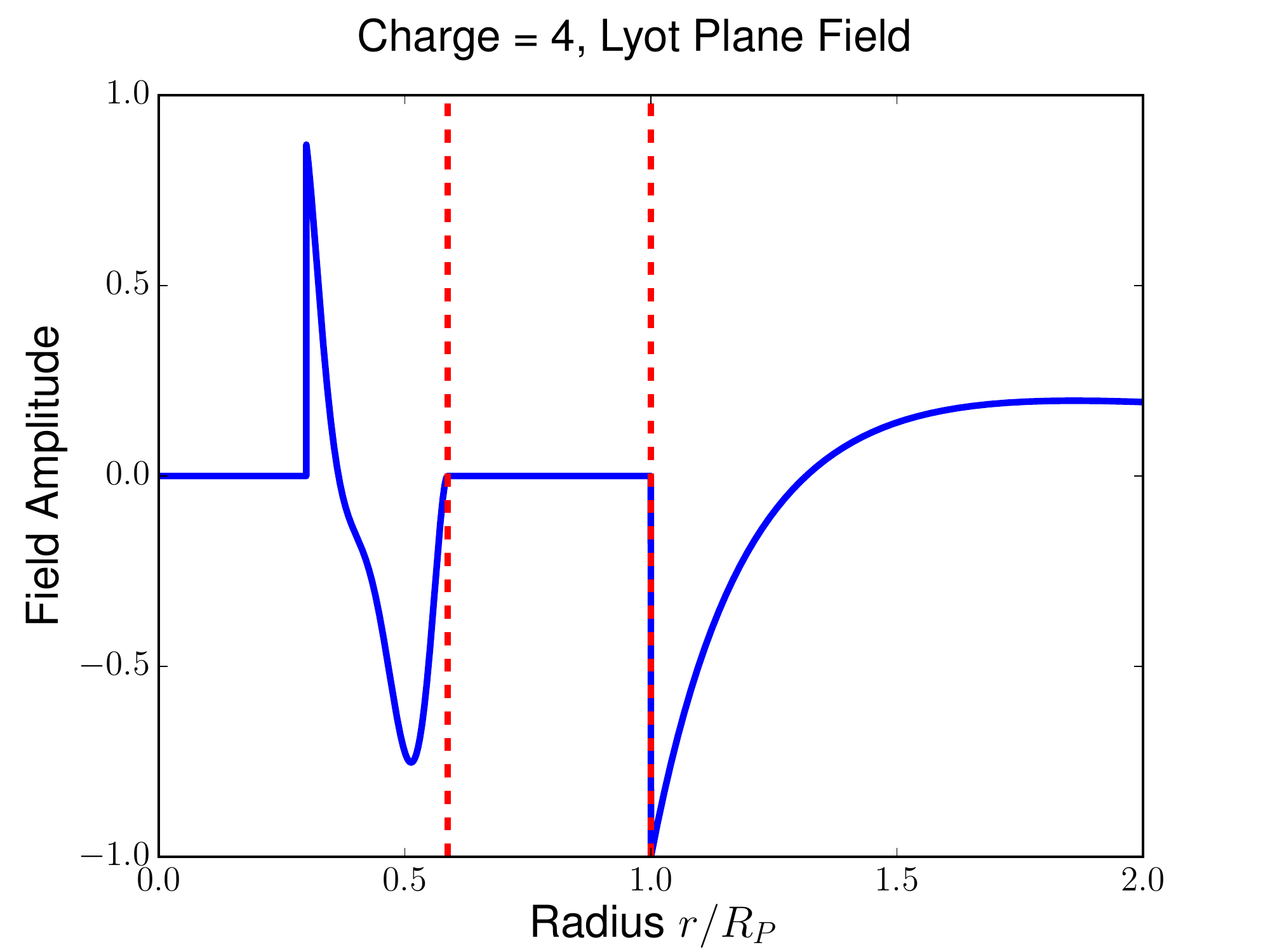}}
\newline
\subfloat{\includegraphics[height=6.5cm]{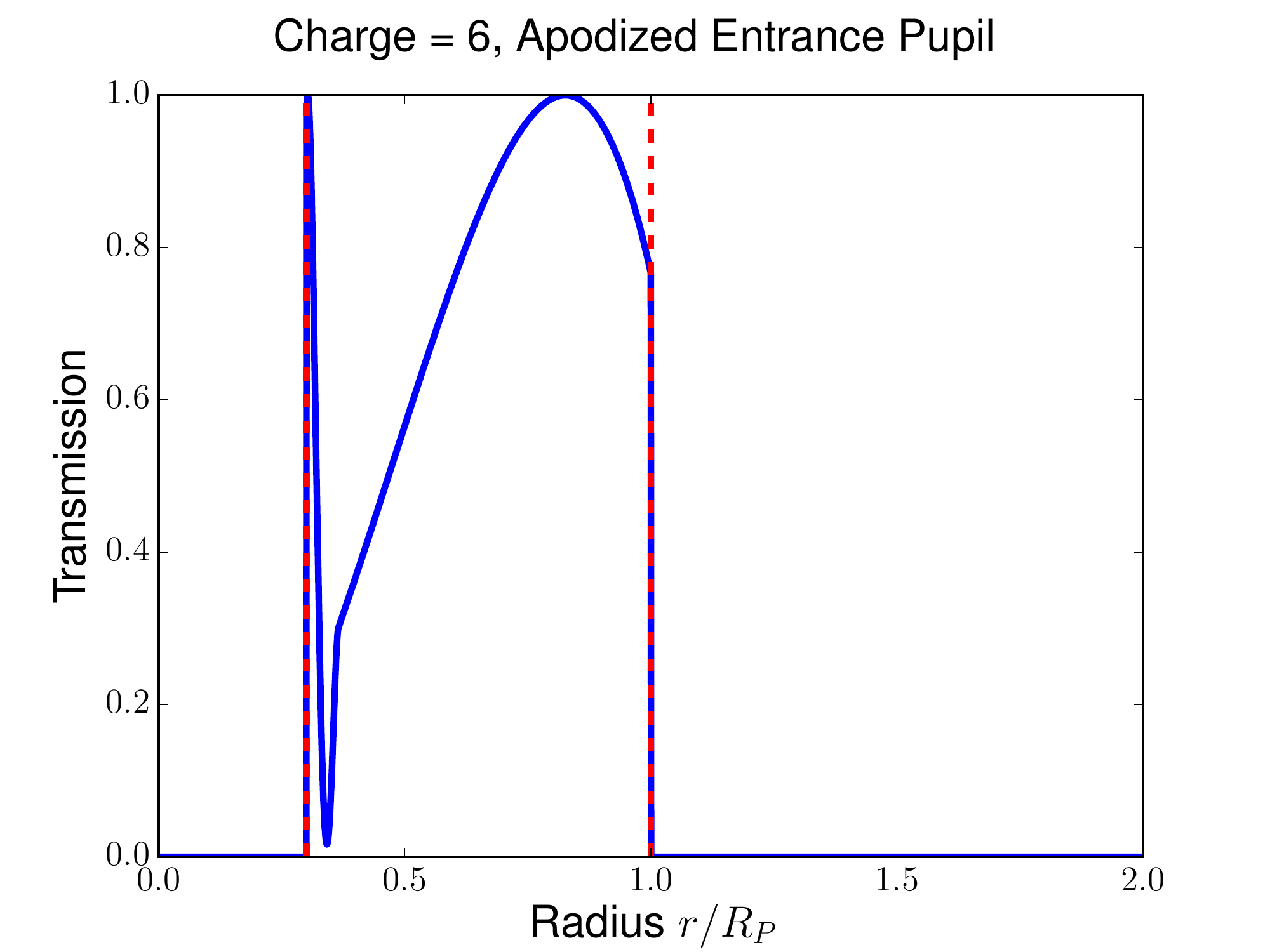}}
~
\subfloat{\includegraphics[height=6.5cm]{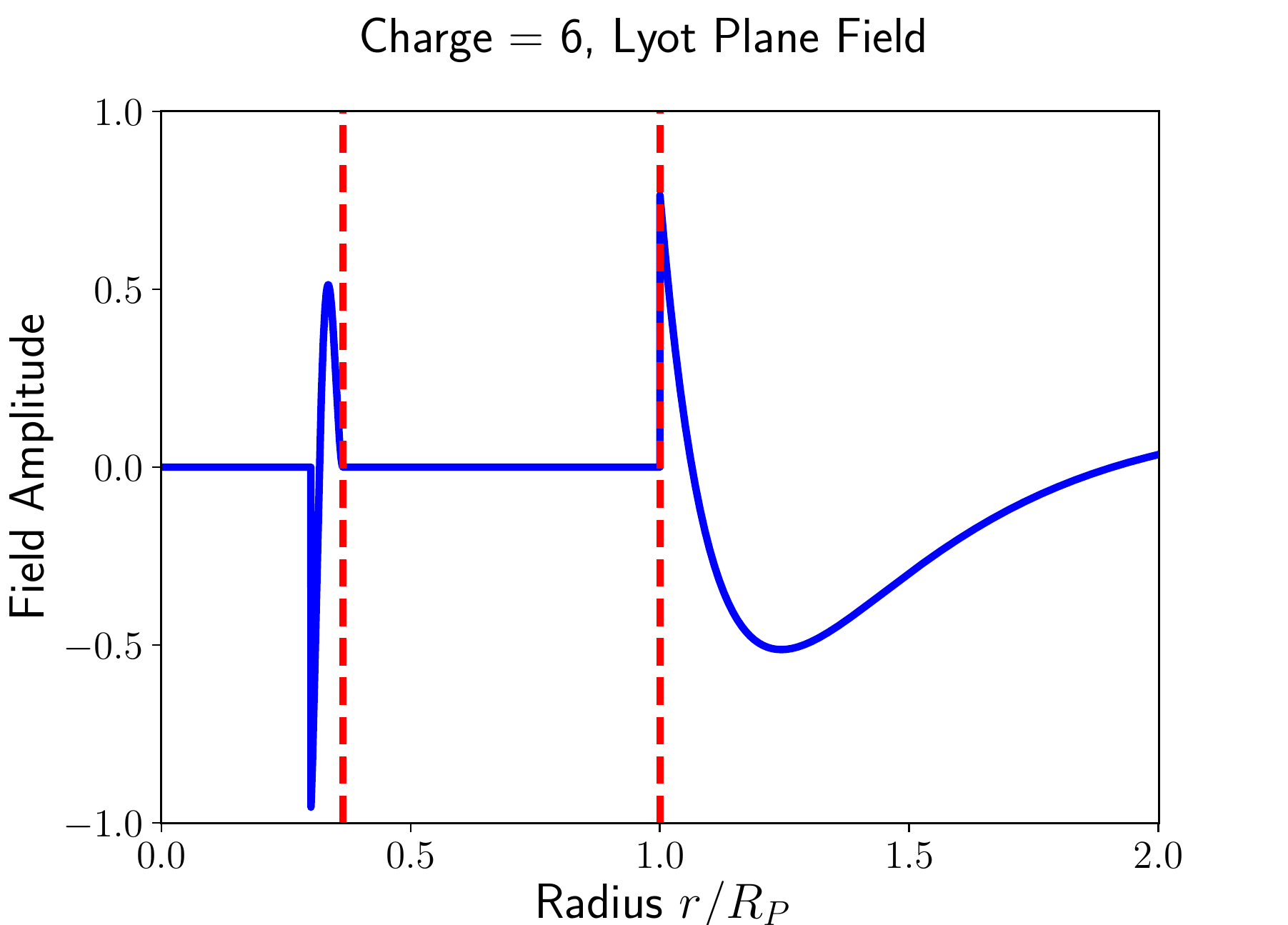}}
\caption[]
{ \label{fig:Optimal_Designs} \textit{Top Row:} The left-hand plot depicts the optimal smooth apodizing mask geometry for a coronagraph designed with a charge 2 vortex for a telescope with $R_{S} = 0.3R_{P}$. Dashed red lines depict the secondary and primary mirror radii. The right-hand panel depicts the Lyot field going out to a radius of twice the radius of the pupil. Dashed red lines depict the inner Lyot stop and pupil radii. \textit{Middle Row:} The same set of plots are shown as in the above row, this time for a coronagraph designed with a charge 4 vortex. \textit{Bottom Row:} The same set of plots are shown as in the above two rows, but for a charge 6 vortex.}
\end{figure*}

In order to explore the PAVC parameter space, we generated optimal smooth apodizing masks for charge 2, 4, and 6 vortices described by various orders of polynomial for different sized central obscurations. We also generated masks with non-continuous apodization functions, where we did not impose the constraints in Equations \ref{eq:10} and \ref{eq:11}. Profiles of smooth apodizing masks with apodization functions described by polynomials of order 20 are shown in Figure \ref{fig:Optimal_Designs} for charge 2 through 6 PAVCs designed for a central obscuration with $R_{S} = 0.3R_{P}$.

\begin{figure*}
\begin{center}
\begin{tabular}{c}
\includegraphics[height=15cm]{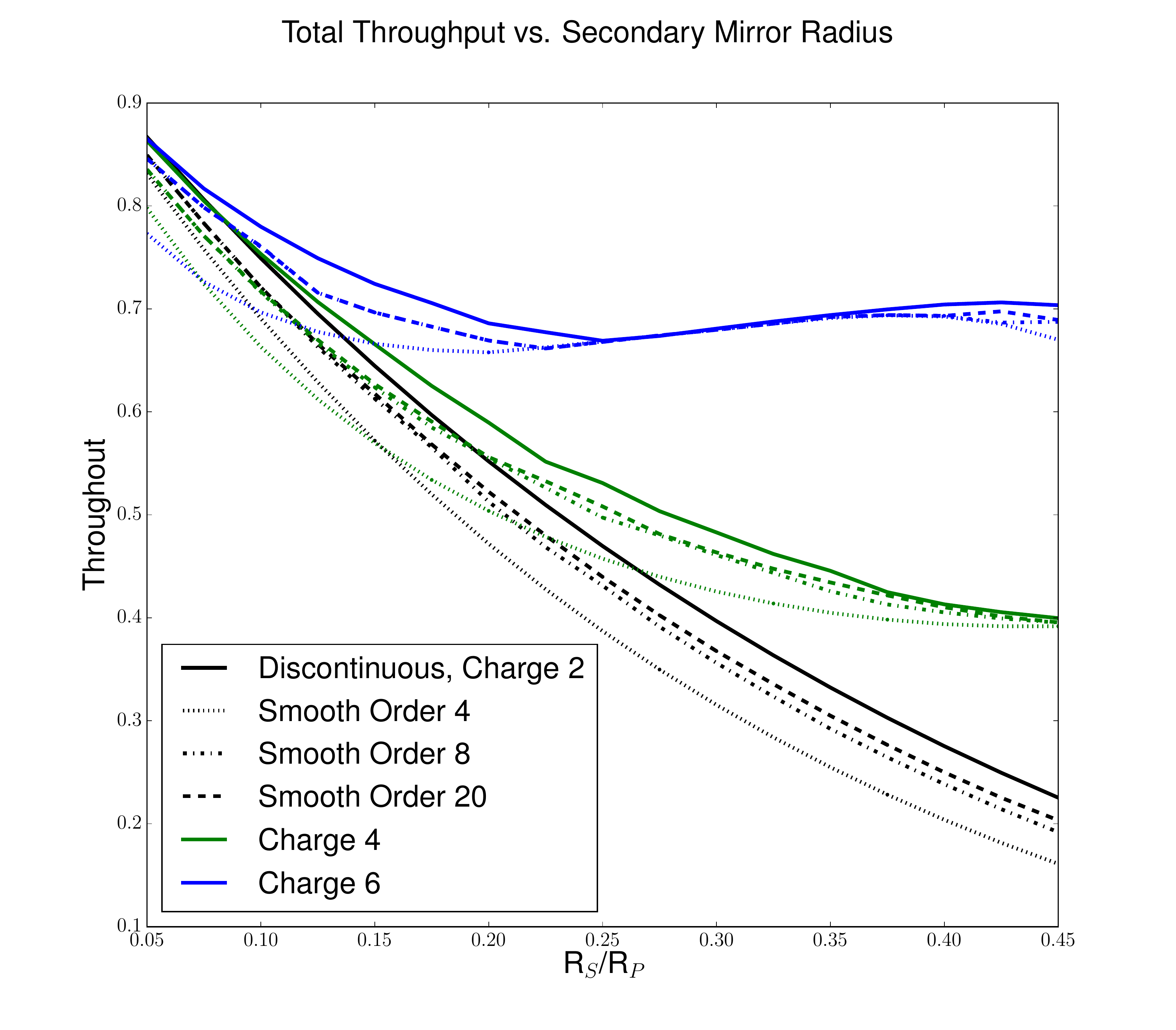}
\end{tabular}
\end{center}
\caption[Throughput vs. Secondary Mirror Radius]
{ \label{fig:TvsRA} Total energy throughput of charge 2, 4, and 6 apodized vortex coronagraphs are shown as a function of the secondary mirror radius relative to the primary mirror radius, where an apodizing filter is used. Solid curves are for apodizations that are not required to be smooth, while dotted, dash-dotted, and dashed lines are for smooth apodizations described by polynomials of order 4, 8, and 20. The black curves are for a $c=2$ coronagraph, and the black solid line corresponds to the ring-apodized vortex coronagraph. The green curves are for a $c=4$ coronagraph, and the blue curves are for a $c=6$ coronagraph.}
\end{figure*}

The optimal total energy throughput achieved by our linear programming algorithm as a function of secondary mirror radius is shown in Figure \ref{fig:TvsRA}. For each $R_{S}$ we chose the value of $R_{I}$ that maximizes $T$. We find that the non-continuous apodizations have the highest total energy throughput for each charge, and that smooth apodizations approach the throughput of the non-continuous apodization for a given charge as the order of the smooth piecewise polynomial increases.

\subsection{Total Energy Throughput Performance}

Since $A\left(r\right)$ need not be continuous for a classically apodized PAVC, the optimal apodization function for charge 2 PAVC is the same as for the charge 2 RAVC. Therefore, the ideal classically apodized charge 2 vortex coronagraph is the RAVC. This is not surprising, since the Lyot plane nulling constraint for a charge 2 PAVC (Equation \ref{eq:C2Null}) constrains the portion of the apodization function where $r>R_{I}$ to be flat. For higher charges, we find apodization functions with substantially higher total energy throughputs than the RAVC apodizations presented in \cite{2013Mawet_RAVC}. In fact, for a given central obscuration size, we find that the total energy throughput of the PAVC increases as the charge increases. This improvement stems from the fact that for higher charges, the apodization function at $r>R_{I}$ can be expressed by progressively higher orders of polynomial. 

The increasing complexity of $A\left(r\right)$ at $r>R_{I}$ with increasing charge in turn explains why total energy throughput vs. $R_{S}$ for the charge 6 PAVC is not perfectly monotonic-- maximizing $T$ may start to noticeably underestimate total energy throughput through the apodizing mask. While $T=\int_{R_{I}}^{R_{P}}{\left[a_{0} + b_{0} + \left(a_{2} + b_{2}\right)r^{2} + \left(a_{4} + b_{4}\right)r^{4}\right]rdr}$, the total energy throughput is $T.E. = \int_{R_{I}}^{R_{P}}{\left[a_{0} + b_{0} + \left(a_{2} + b_{2}\right)r^{2} + \left(a_{4} + b_{4}\right)r^{4}\right]^{2}rdr}$, which has an integrand with higher-ordered terms and three additional cross terms. Meanwhile, optimizing $T$ for a charge 4 vortex provides a better approximation for optimizing throughput, since there are fewer terms in both the integrands for $T$ and $T.E.$, and fewer cross terms in the integrand for throughput. For both the charge 4 and 6 PAVC, the theoretically optimal total energy throughput is slightly higher than what we report. However, for the purposes of demonstrating the characteristics of the PAVC, the approximate optimal total energy throughput curves we obtain by optimizing $T$ suffice.

\begin{figure*}
\begin{center}
\begin{tabular}{c}
\includegraphics[height=12cm]{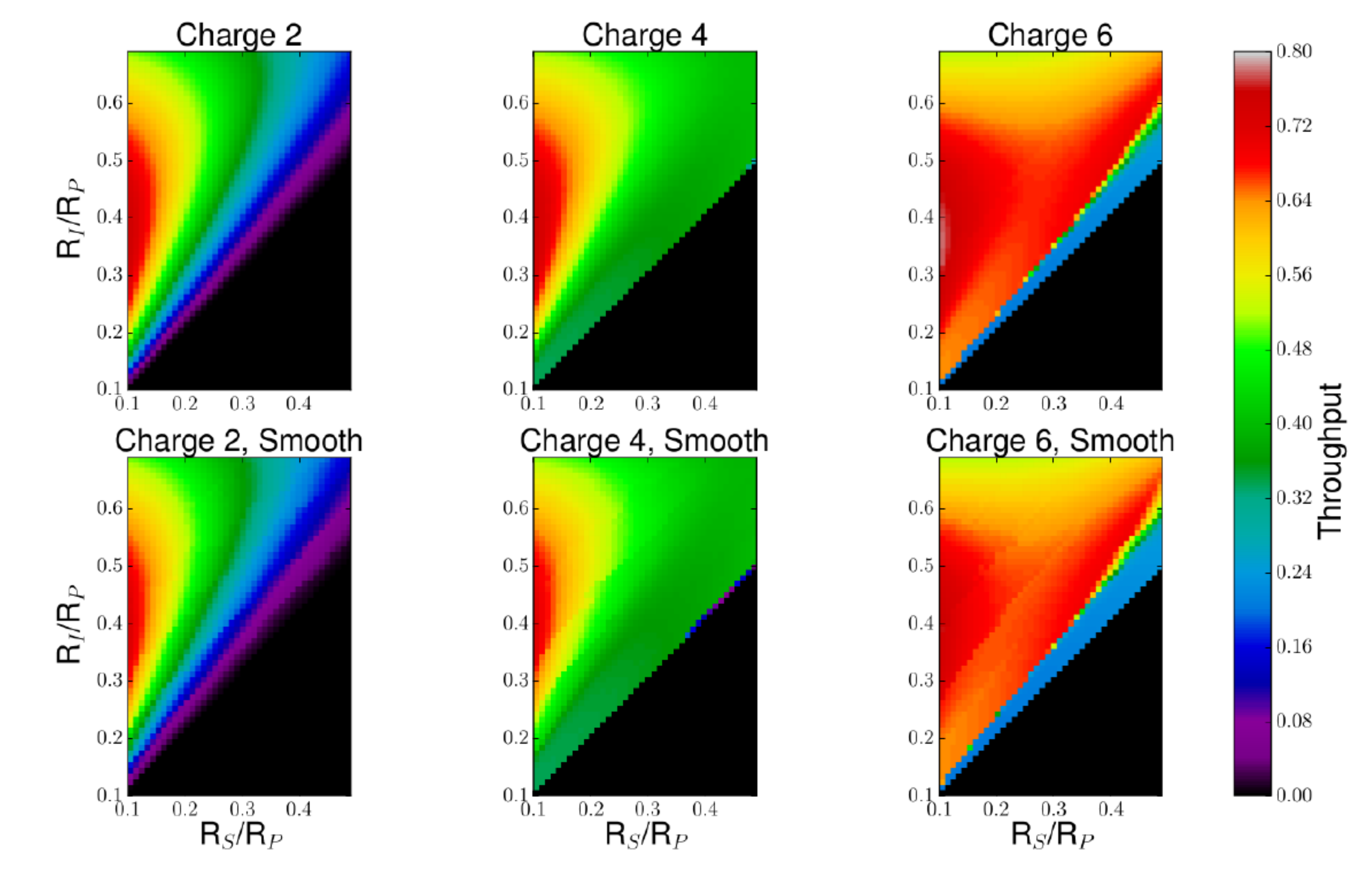}
\end{tabular}
\end{center}
\caption[Inner Lyot Stop Radius vs. Secondary Mirror Radius]
{ \label{fig:RBvsRA} \textit{Top Row:} Total energy throughputs of apodizing masks optimized to maximize transmission for combinations of inner Lyot stop radius $R_{I}$ and secondary mirror radius $R_{S}$ for charge 2, 4, and 6 vortex phase masks. Results are shown for discontinuous apodizations. $R_{I}$ and $R_{S}$ are shown in units relative to the primary mirror radius, $R_{P}$. Throughput for coronagraph designs where $R_{S} > R_{I}$ are 0. \textit{Bottom Row:} Results for smooth apodizations described by order-20 polynomials.}
\end{figure*}

Total energy throughput as a function of both $R_{S}$ and $R_{I}$ is shown in Figure \ref{fig:RBvsRA}. We find that the total energy throughput as a function of either radius increases with charge. For the charge 2 PAVC, the optimal value of $R_{I}$ increases monotonically with $R_{S}$, while the dependence of the optimal value of $R_{I}$ on $R_{S}$ is more complicated for the charge 4 and 6 cases. Indeed, the throughput surface for the charge 6 coronagraph exhibits a saddle point at around $R_{S}/R_{P} = 0.3$ and $R_{I}/R_{P} = 0.4$, owing to the competing effects of throughput loss due to increased Lyot stop size and gain in the apodizing mask transmission. Like the result that PAVC total energy throughput improves as charge increases, the behavior of the charge 4 and 6 throughput surfaces results from the increasing number of available degrees of freedom at $r>R_{I}$ as the charge increases. This is true whether or not the PAVC apodization is constrained to be smooth. As shown in Figure \ref{fig:RBvsRA}, forcing the apodization to be smooth affects the performance of the PAVC, but throughputs for both smooth and discontinuous apodizations are similar and depend on $R_{S}$ and $R_{I}$ in nearly the same way.

\subsection{Off-Axis Source Performance}

\begin{figure*}
{\centering
\subfloat{\includegraphics[height=6.125cm]{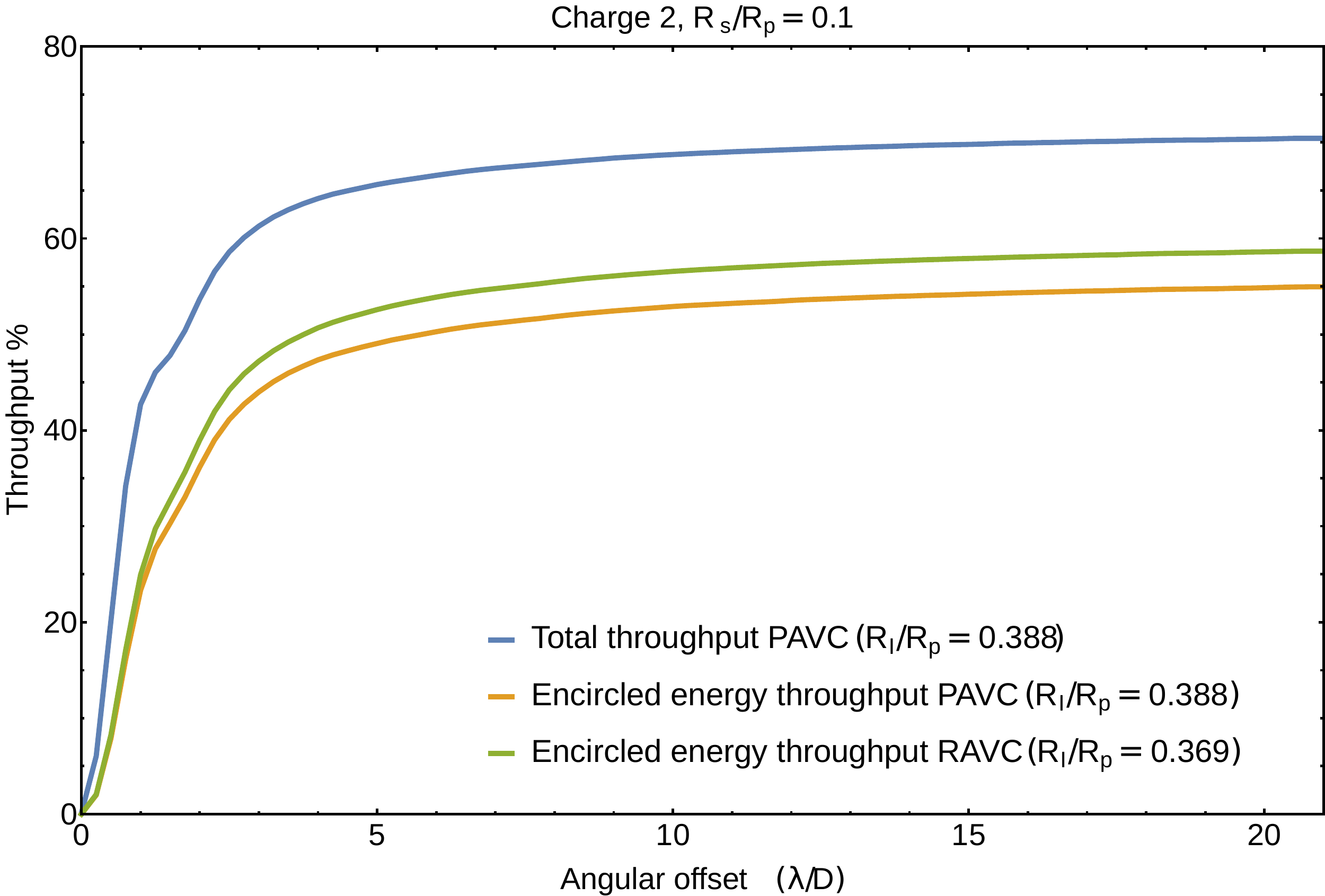}}
~
\subfloat{\includegraphics[height=6.125cm]{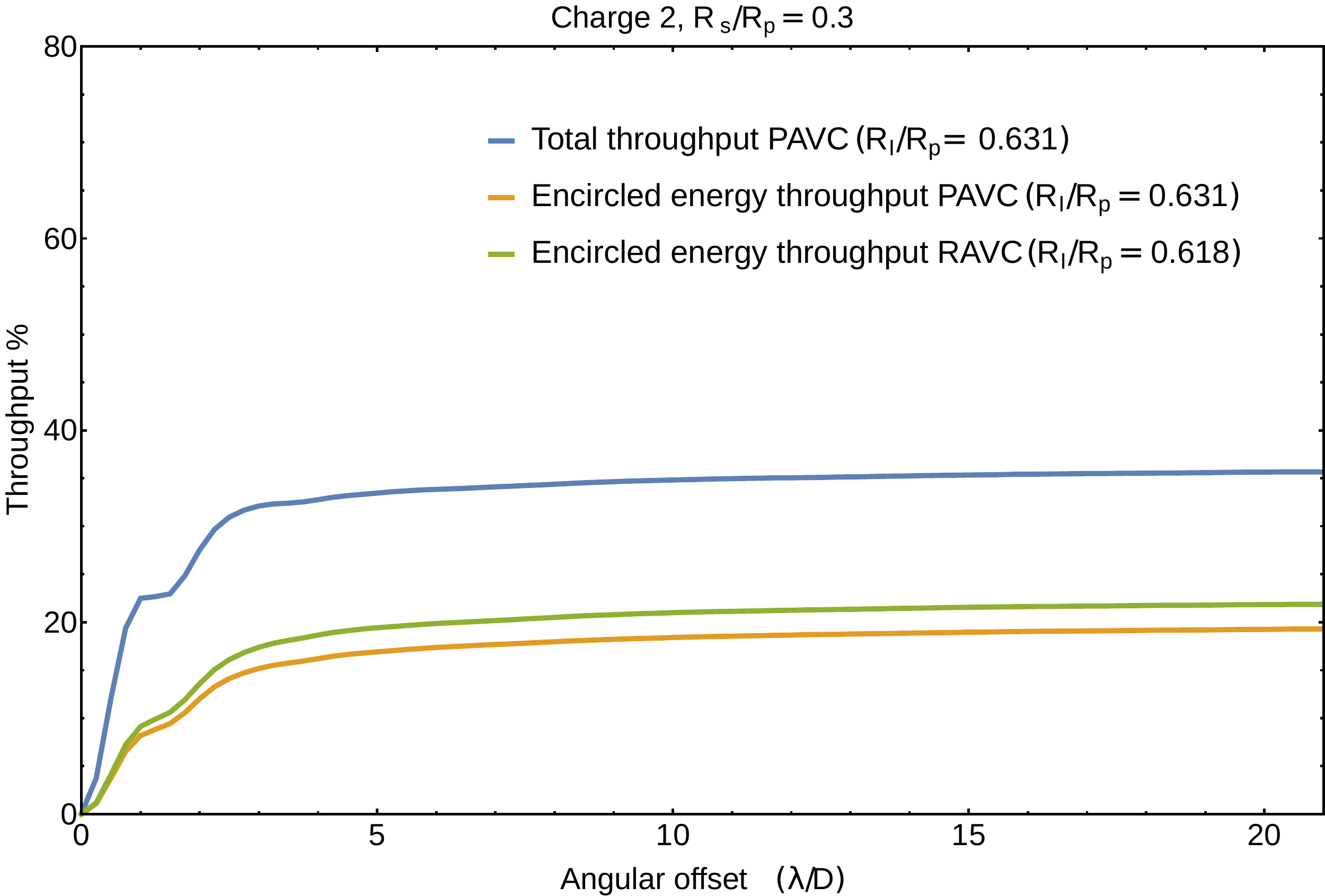}}
\newline
\subfloat{\includegraphics[height=6.125cm]{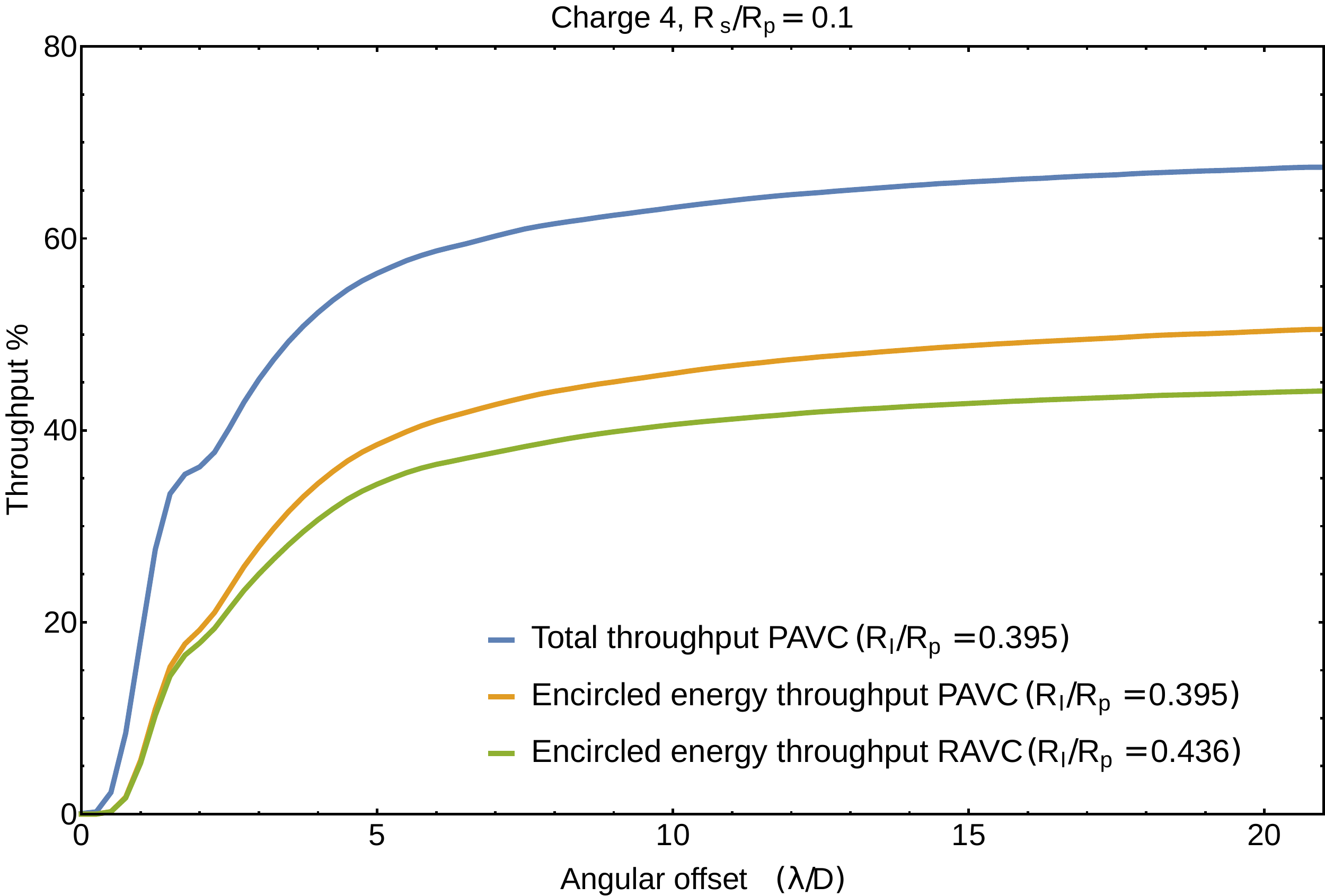}}
~
\subfloat{\includegraphics[height=6.125cm]{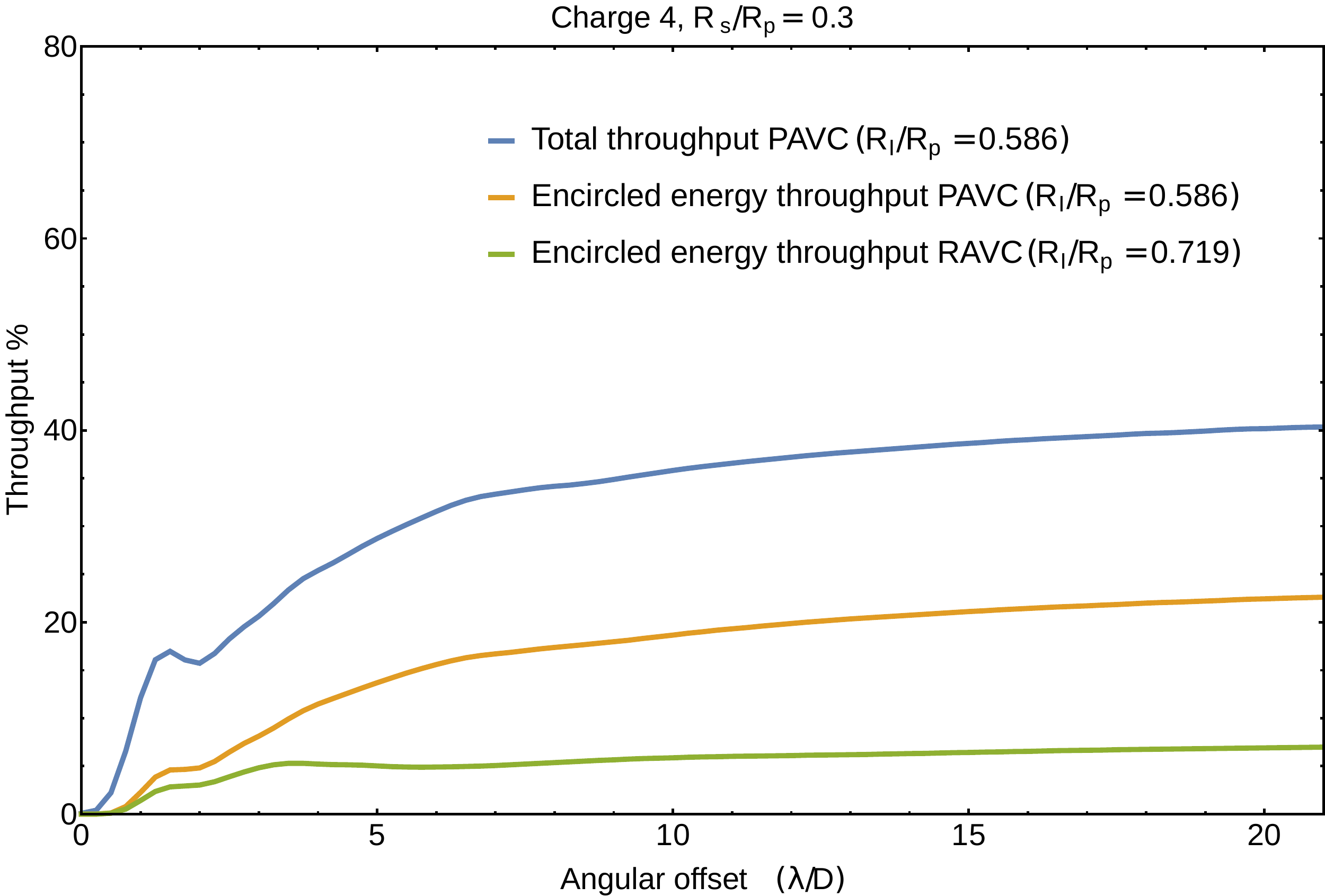}}
\newline
\subfloat{\includegraphics[height=6.125cm]{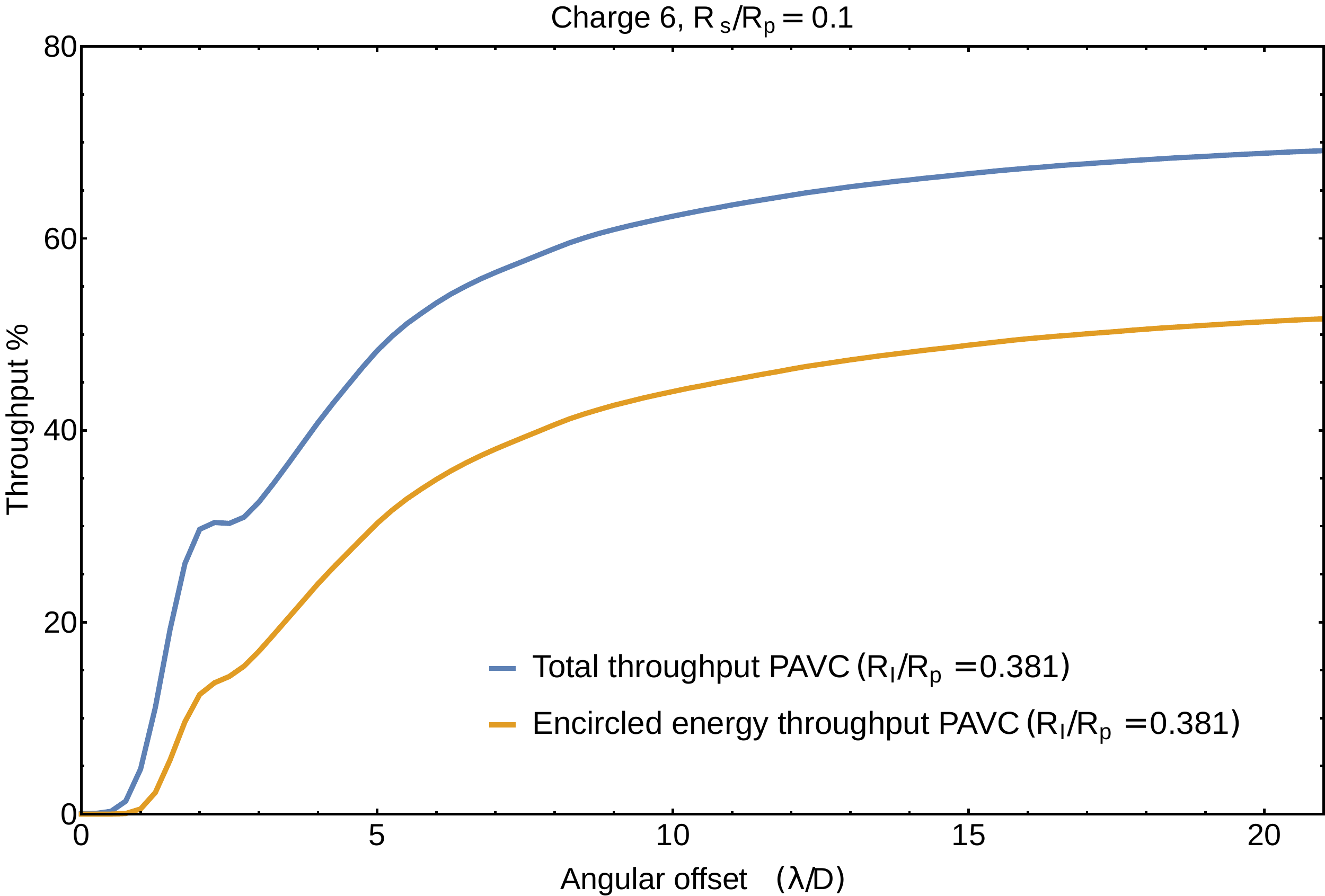}}
~
\subfloat{\includegraphics[height=6.125cm]{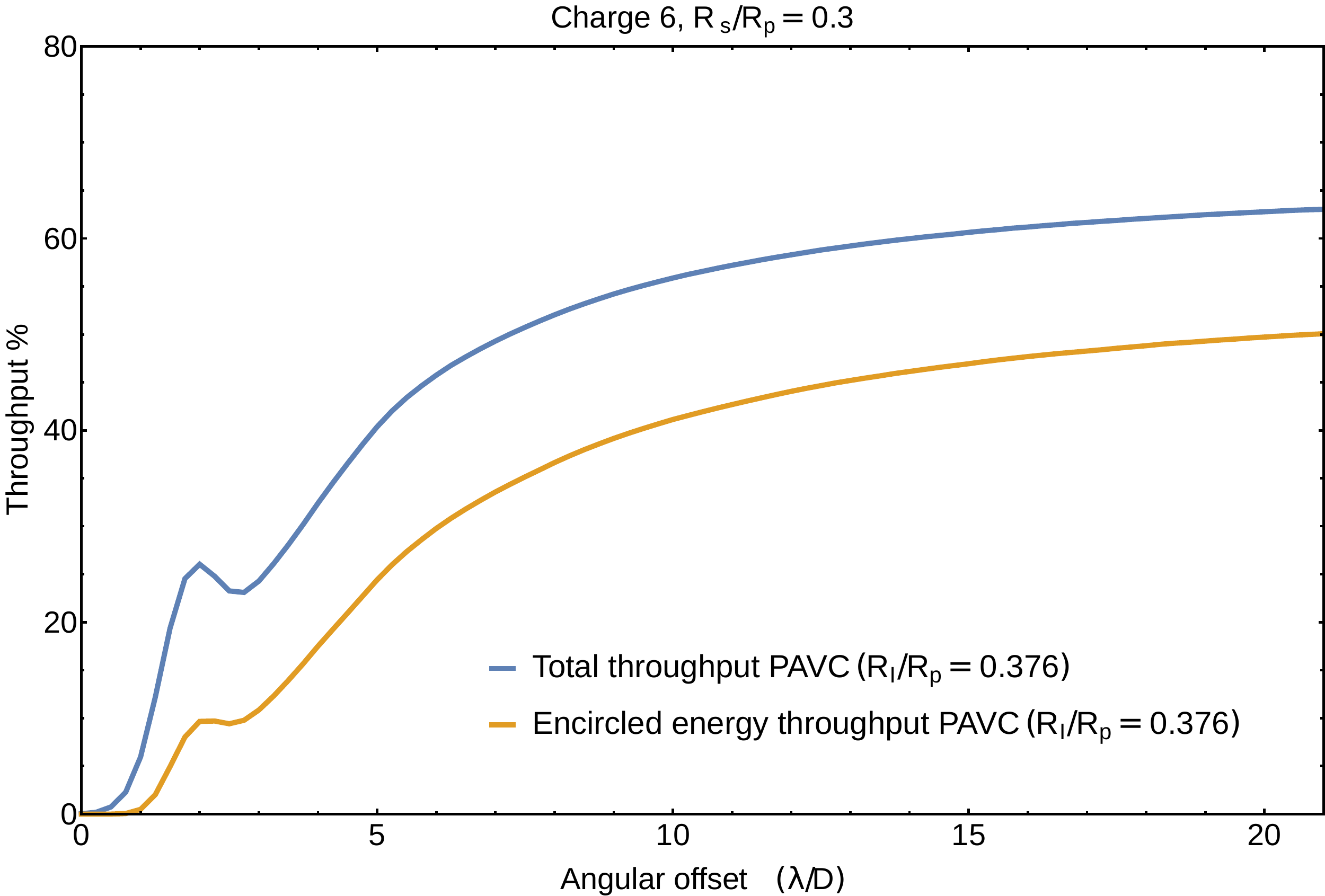}}}

\caption[Mirror Shapes]
{ \label{fig:Off_Axis} \textit{First Row:} Off-axis throughputs for optimal coronagraph designs with a charge 2 vortex. The left-hand panel shows the throughput of an off-axis source, such as a planet, as a function of distance from the on-axis star in units of $\lambda$/D for a coronagraph designed for a telescope with $R_{S} = 0.1R_{P}$ The blue line is the total throughput, and the yellow line is the encircled energy throughput. For comparison, the green line shows the encircled energy throughput of the charge 2 RAVC for the same secondary mirror radius. The right-hand panel show the same set of curves, this time for a coronagraph designed for a telescope with $R_{S} = 0.3R_{P}$. \textit{Middle Row:} The same set of plots as the top row, but for a coronagraph designed with a charge 4 vortex. \textit{Bottom Row:} The same set of plots as the top two rows, but for a coronagraph designed with a charge 6 vortex. An RAVC comparison curve is not shown, since the charge 6 RAVC was not considered in \cite{2013Mawet_RAVC}, owing to its low throughput.}
\end{figure*}

Total and encircled energy throughputs for an off-axis source (see Table \ref{table:Throughput_Defs}) as a function of the source's angular separation from the on-axis star are shown in Figure \ref{fig:Off_Axis}. We present curves depicting throughput vs. angular separation for charge 2, 4, and 6 PAVCs, and for $R_{S} = 0.1R_{P}$ and $R_{S} = 0.3R_{P}$. By plotting the encircled energy throughout, we provide an estimate of the actual flux an observer would measure when performing aperture photometry on the final coronagraphic image \citep{2016Ruane_Apod}. Plotting the total energy throughput as well indicates the total amount of flux from the off-axis source present in the final coronagraphic image. 

In the charge 2 and 4 plots, the encircled energy throughput curve of the corresponding charge RAVC is shown for reference. The smooth charge 2 PAVC underperforms the charge 2 RAVC in terms of throughput, since the ideal charge 2 apodized vortex coronagraph is the RAVC. However, the situation is the reverse for the charge 4 PAVC, especially when $R_{S} = 0.3R_{P}$. In that case the smooth PAVC outperforms the off-axis throughput of the RAVC by a factor of 3. We do not show RAVC curves on our charge 6 plots, since in \cite{2013Mawet_RAVC}, charge $> 4$ RAVCs were not discussed due to low total energy throughput performance. 

 While going from an 0.1$R_{P}$ central obscuration to an 0.3$R_{P}$ central obscuration reduces the encircled energy throughput for an off-axis source by a little more than half for the charge 4 PAVC (from $\sim 50\%$ to $\sim 22\%$), for the charge 6 PAVC the maximum off-axis throughput is essentially unchanged (changing from $\sim 52\%$ to $\sim 50\%$). However, when going from a central obscuration of 0.1$R_{P}$ to 0.3$R_{P}$, the angular separation at which the encircled energy throughput reaches $20\%$ increases from $\sim 3.5$ $\lambda$/D to $\sim 4.25$ $\lambda$/D for the charge 6 PAVC. By comparsion, the charge 4 PAVC with a central obscuration of 0.1$R_{P}$ obtains a 20$\%$ encircled energy throughput at an angular separation of $\sim 2.0$ $\lambda$/D. 

The charge 6 PAVC addresses the issue of throughput performance for coronagraphs designed for telescopes with large secondary mirrors. The charge 4 PAVC, meanwhile, represents a trade-off between sensitivity to central obscuration size and inner working angle. 

Either the charge 4 or charge 6 PAVC may provide the optimal basis for designing an instrument for an on-axis telescope. Depending on the secondary mirror size, as well as the relative importance of overall throughput performance and the minimum angular separation for an observable off-axis source, one version of the PAVC may be preferable over the other. The charge 2 RAVC may be preferable on ground-based telescopes with relatively small secondary mirrors when sensitivity to low-order aberrations is not a priority but a small IWA is.

\section{The Shaped Mirror PAVC}

\begin{figure}
\begin{center}
\includegraphics[height=4.0cm]{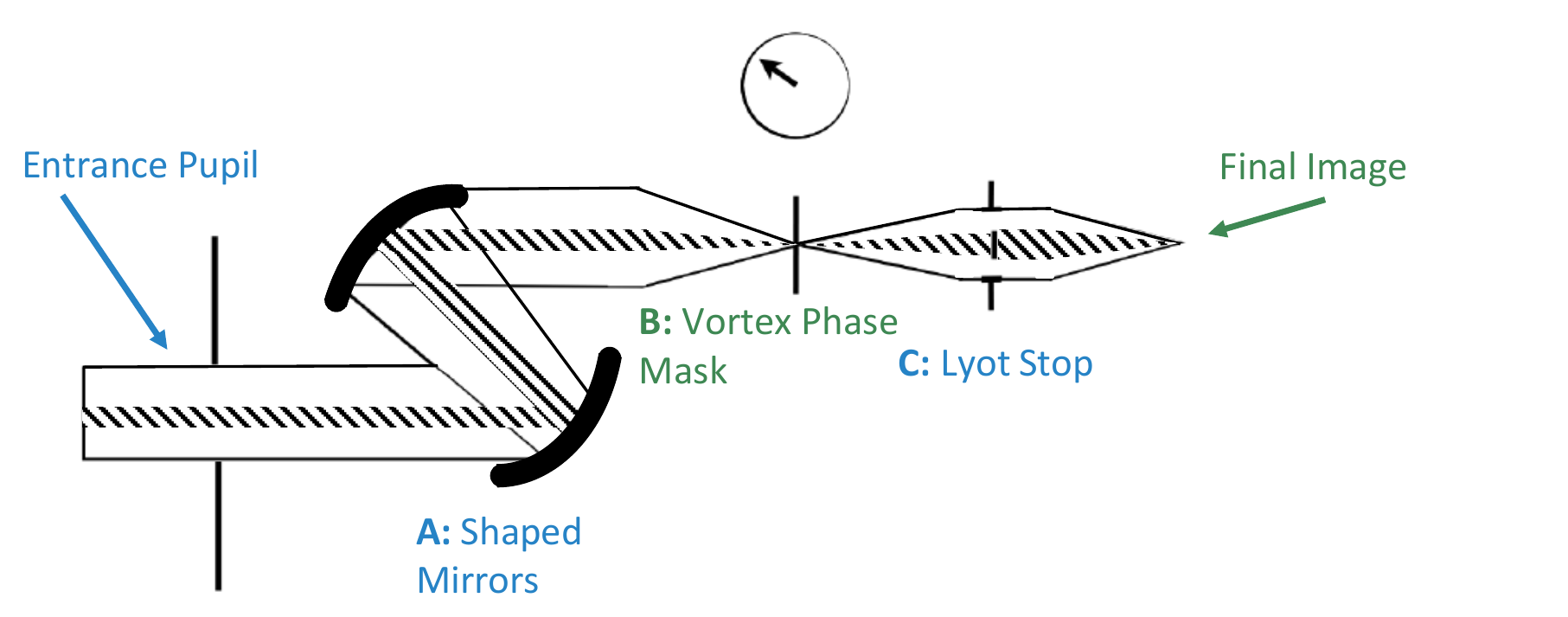}
\end{center}
\caption[]
{ \label{fig:Mirror_Overview} The coronagraph design from Figure \ref{fig:Overview} modified to replace the pupil plane apodizer with a pair of shaped mirrors. The shaped mirrors remap the on-axis flux to has the pupil geometry necessary to null the Lyot field. As in Figure \ref{fig:Overview}, parts of the beam labeled in blue are in the pupil plane and those in green are in the focal plane.}
\end{figure}

We now show how the PAVC can be implemented with shaped mirrors. In essence, we are combining the vortex coronagraph with PIAA shaped mirror apodization, except that the shapes of the mirrors and apodizations we use differ significantly from those in \cite{2014Guyon}. We modify the PAVC setup slightly, replacing the apodizing filter in Figure \ref{fig:Overview} with a pair of shaped mirrors, as shown in Figure \ref{fig:Mirror_Overview}. In some cases it may be advantageous to reshape the pupil with mirrors, instead of apodizing with a filter, since there is no loss of flux due to transmission through the filter, and total energy throughput is limited by the size of the Lyot stop. We first calculate the mirror shapes needed for PAVC apodization and describe the tradeoff between the total energy throughput and mirror curvature. Then, we find mirror shapes that produce apodizations for a charge 4 and charge 6 PAVC.

\subsection{Mirror Apodizations for the PAVC}

We obtain expressions to describe the circularly symmetric mirror shapes we use to construct an apodizing shaped pupil following the derivation given in \cite{2005Vanderbei} (V05), with some modifications. A pair of mirrors is used to produce a desired apodization function-- the first mirror `spreads out' the incoming point source flux, and the second mirror re-collimates the beam. We describe the shape of the first mirror with coordinates $\left(\rho, \theta\right)$ and mirror height (relative to a flat mirror) $h\left(\rho,\theta\right)$. The second mirror is in the pupil plane, so its coordinates are  $\left(r,\theta\right)$ and its height is $\tilde{h}\left(r,\theta\right)$.  We drop the $\theta$ dependence throughout, since the mirror shapes and apodization functions we wish to solve for are circularly symmetric, and note that the more general expressions may be found in V05.

To solve for the mirror heights, we use Equations 24, 29\footnote{There is a sign error in equation 29 of V05, which reads $H\left(\rho, \theta\right) = \tilde{H}\left(r,\theta\right)-\frac{P_{0}}{2}-\frac{\left(r-\rho\right)^{2} + 2\left(r-\rho\right)\delta \cos\left(\theta\right)+\delta^{2}}{2P_{0}}$, but ought to read $H\left(\rho, \theta \right) = \tilde{H}\left(r,\theta\right)-\frac{P_{0}}{2}+\frac{\left(r-\rho\right)^{2} + 2\left(r-\rho\right)\delta \cos\left(\theta\right)+\delta^{2}}{2P_{0}}$. Note that V05 uses coordinates $\left(r, \theta\right)$, $\left(\tilde{r}, \theta\right)$ instead of $\left(\rho, \theta\right)$, $\left(r, \theta\right)$; we express their equation 29 in the coordinates we use in this paper for clarity.} and 34 in V05, which, dropping the $\theta$ dependence and other constants, are 
\begin{equation}\label{eq:18}
\partial_{r}\tilde{h}  = \rho\left(r\right) - r,
\end{equation}
\begin{equation}\label{eq:19}
h\left(\rho\right) = \tilde{h}\left(r\right) + \frac{1}{2}\left(r-\rho\right)^{2},
\end{equation}
and
\begin{equation}\label{eq:20}
\frac{d}{dr}\rho\left(r\right)^{2} = 2CA^{2}\left(r\right)r.
\end{equation}
The extra coefficient $C$ in Equation \ref{eq:20} normalizes the apodization function, so that energy is conserved across the two mirrors, such that
\begin{equation}\label{eq:21}
2\pi C\int_{0}^{R_{P}}{A^{2}\left(r\right)rdr} = 2\pi R_{P}^{2}.
\end{equation}
In order to solve Equation \ref{eq:20}, we set $A\left(r\right) = 1$ for $r < R_{S} - \epsilon$, and use spline interpolation to smoothly connect $A\left(R_{s}\right)$ and $A\left(R_{S}-\epsilon\right)$, where $\epsilon$ is an arbitrary small separation. We note that it can be shown from the expression for $\rho\left(r\right)$ that the equation for the derivative of the second mirror is equivalent to the second-order Monge-Amp\`ere equation for the second mirror shape presented in \cite{2011Pueyo_Derivs} and \cite{ACAD}.

\begin{figure}[b!]
\begin{center}
\begin{tabular}{c}
\includegraphics[height=8cm]{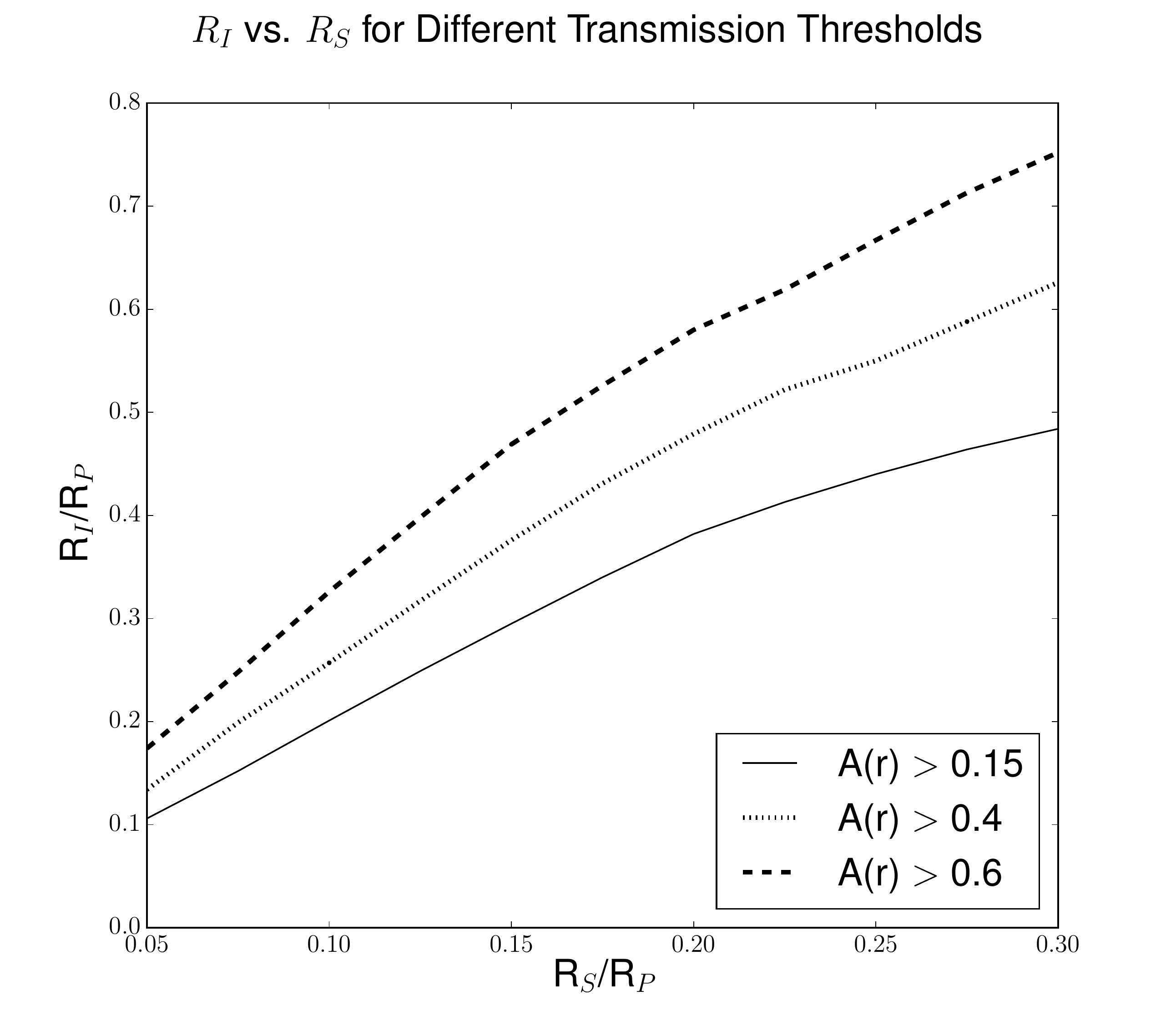}
\end{tabular}
\end{center}
\caption[Inner Lyot Stop Radius vs. Secondary Mirror Radius]
{ \label{fig:RBforMinvsRA} Minimum inner Lyot stop radii allowing for minimum values of $A\left(r\right)$ of 0.15 (solid lines), 0.4 (dotted), and 0.6 (dashed) are shown as a function of secondary mirror radius. For simplicity of the figure, only curves for charge 4 are shown. }
\end{figure}

We wish the optimize $A\left(r\right)$ to maximize the shaped mirrors' radii of curvature. Mirror shapes with sharper curvatures are both more difficult to manufacture and increase the magnitude of diffraction effects introduced by the optics, so we want to use the smoothest mirror shapes possible \citep{2006Pluzhnik_PIAA}. We focus on the shape of the first mirror, which we empirically observe has the sharpest curvatures. The radius of curvature (reciprocal of the curvature, $K$) for this mirror is
\begin{equation}\label{eq:24}
\frac{1}{K} = \frac{\left(1+\left(\partial_{\rho}h\right)^{2}\right)^{3/2}}{\left|\partial^{2}_{\rho}h\right|}.
\end{equation}
Finding $\partial_{\rho}h$ and $\partial^{2}_{\rho}h$ is difficult using the setup described in V05, so we solve for the setup in reverse, where an initial pupil with electric field amplitude $A\left(r\right)$ is reshaped by a series of mirrors to an outgoing beam of uniform amplitude. In this setup, $h\left(r\right)$ is the height of the second mirror, and the desired apodization function is $A^{-1}\left(r\right)$. Making these adjustments to Equations \ref{eq:18} and \ref{eq:19}, we find that
\begin{equation}\label{eq:25}
\frac{1}{K} = \frac{A^{2}\left(1+\left(\frac{2}{A}-r\right)^{2}\right)^{3/2}}{2\partial_{r}A}.
\end{equation}
To maximize the radius of curvature across the mirror shape, it follows that we want to maximize $A$ and minimize $\partial_{r}A$. However, as seen in Figure \ref{fig:RBforMinvsRA}, the minimum value of $A\left(r\right)$ depends on $R_{I}$, which in turn limits the total energy throughput. For a desired level of throughput, then, we can optimize for curvature by adopting $\partial_{r}A$ as the FOM to minimize.

\subsection{Mirror Apodized PAVC Performance}

\begin{figure*}
\centering
\subfloat{\includegraphics[height=6.5cm]{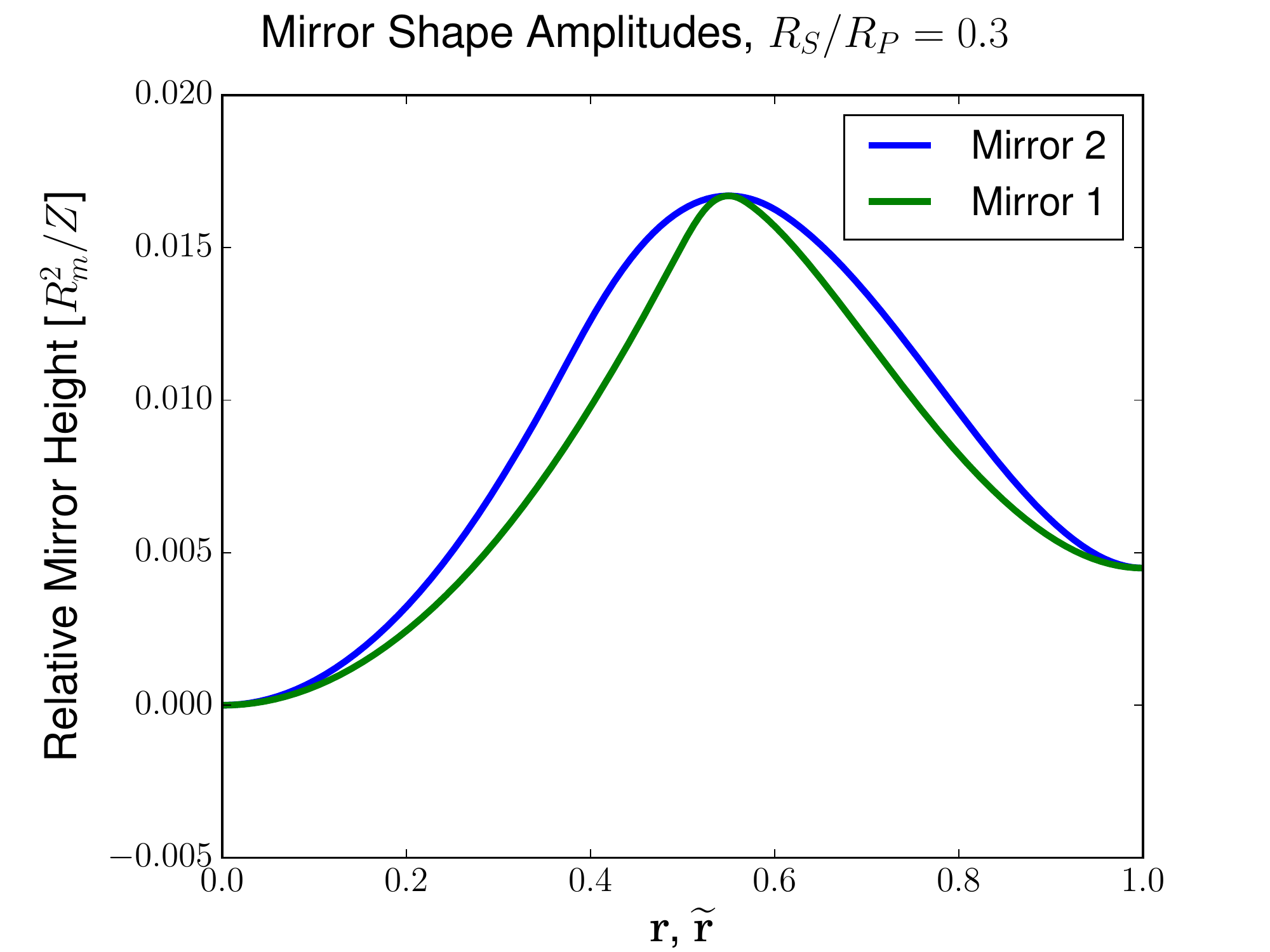}}
~
\subfloat{\includegraphics[height=6.5cm]{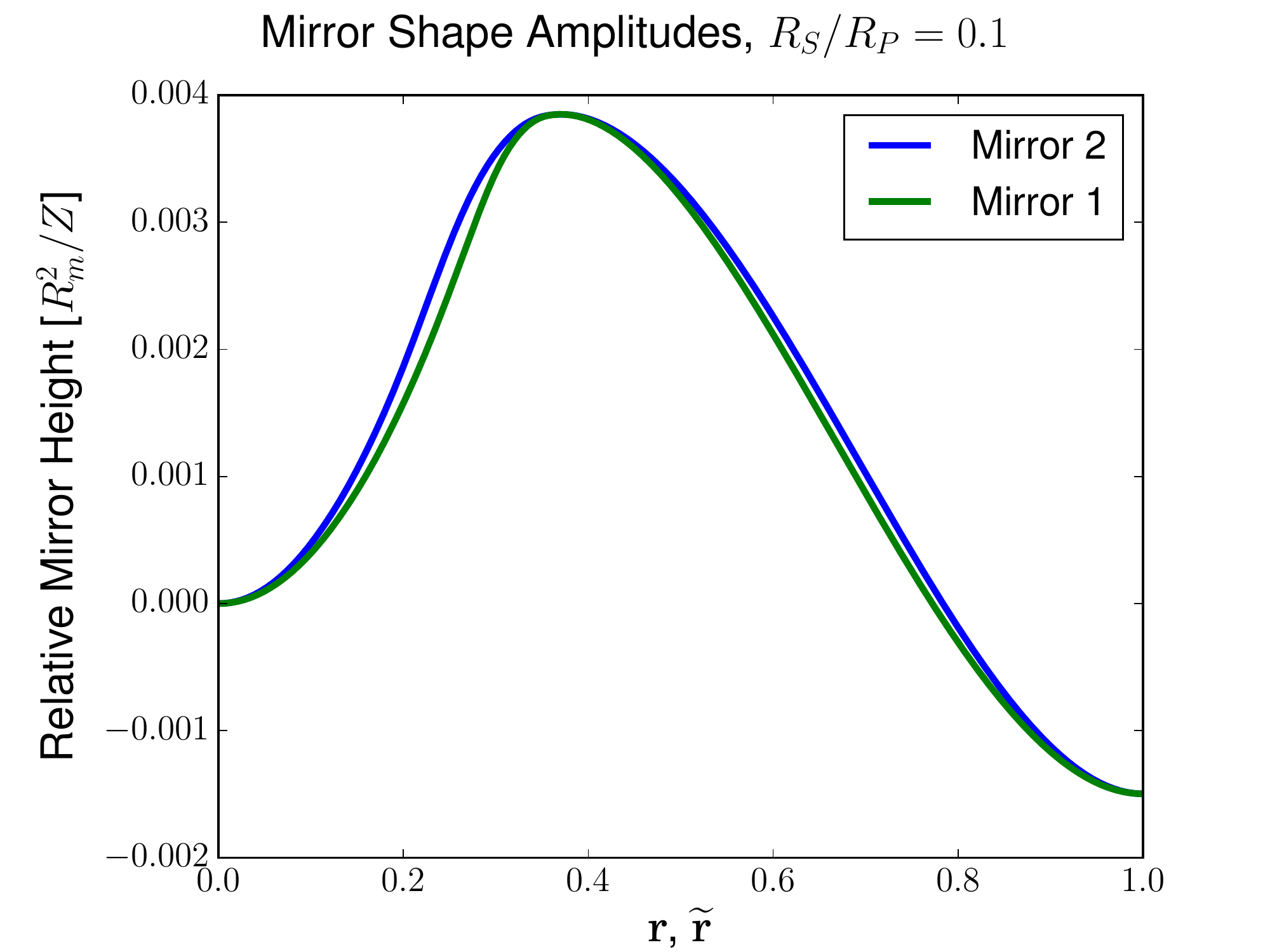}}
\newline
\subfloat{\includegraphics[height=6.5cm]{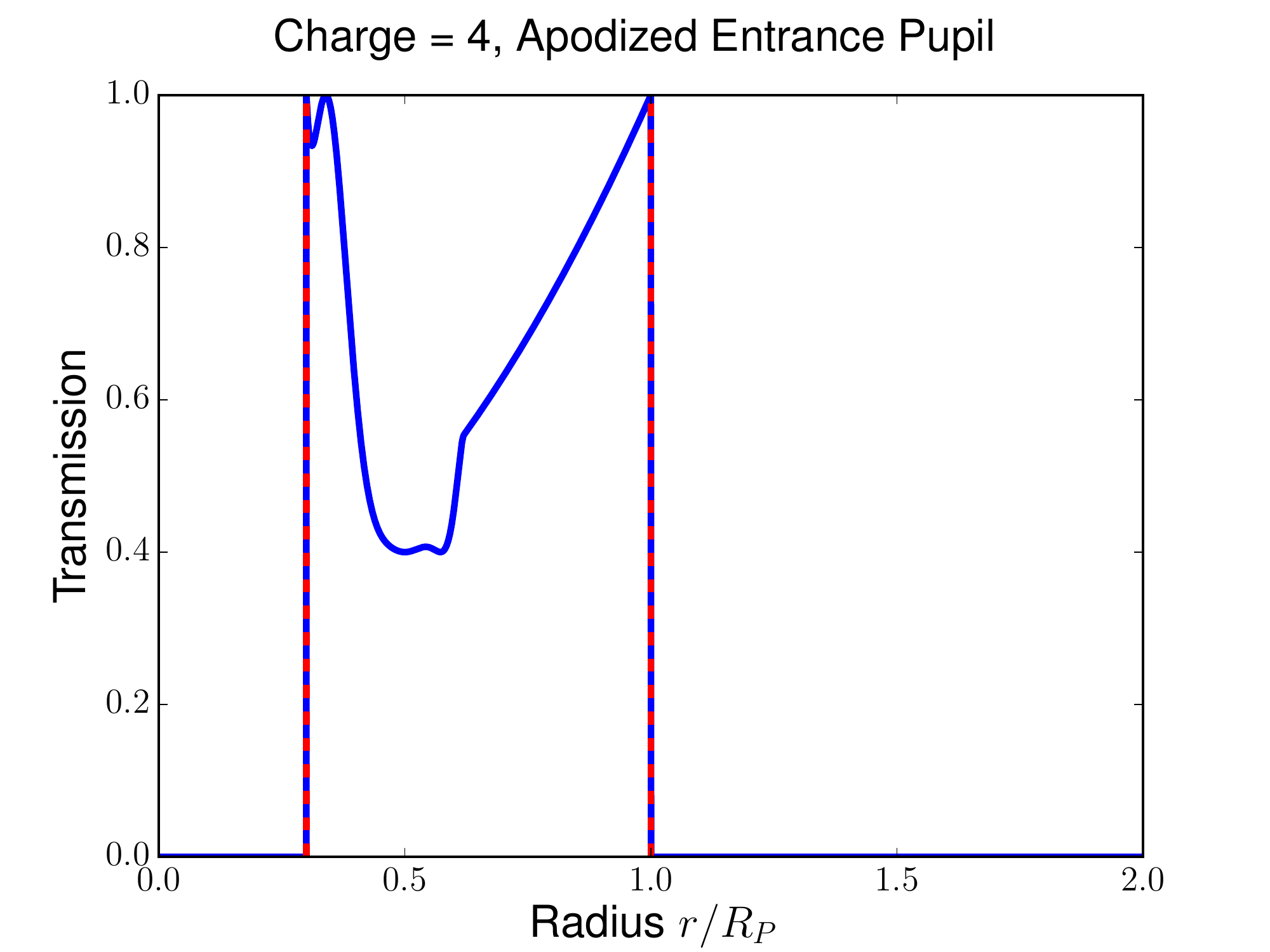}}
~
\subfloat{\includegraphics[height=6.5cm]{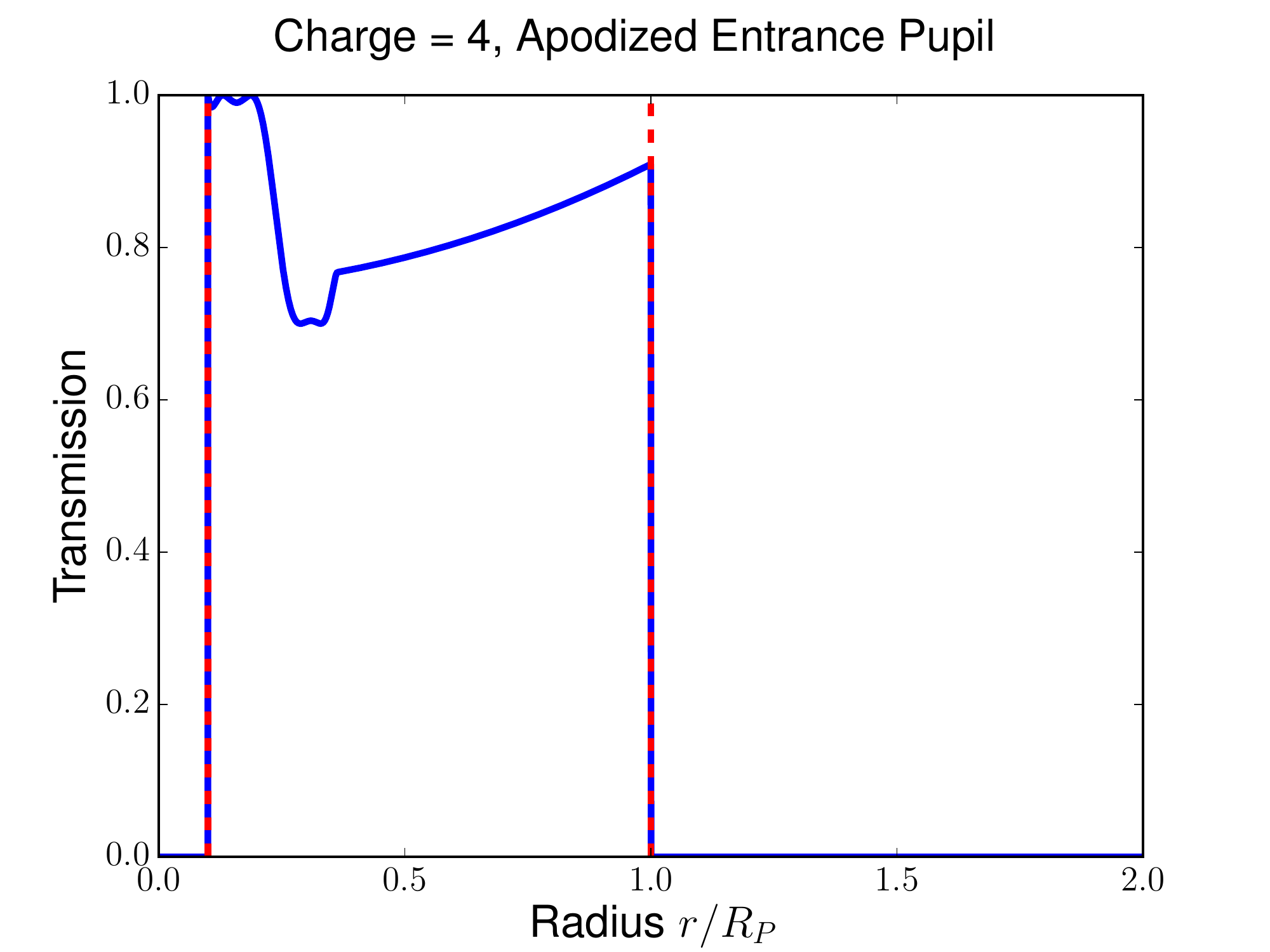}}
\newline
\subfloat{\includegraphics[height=6.5cm]{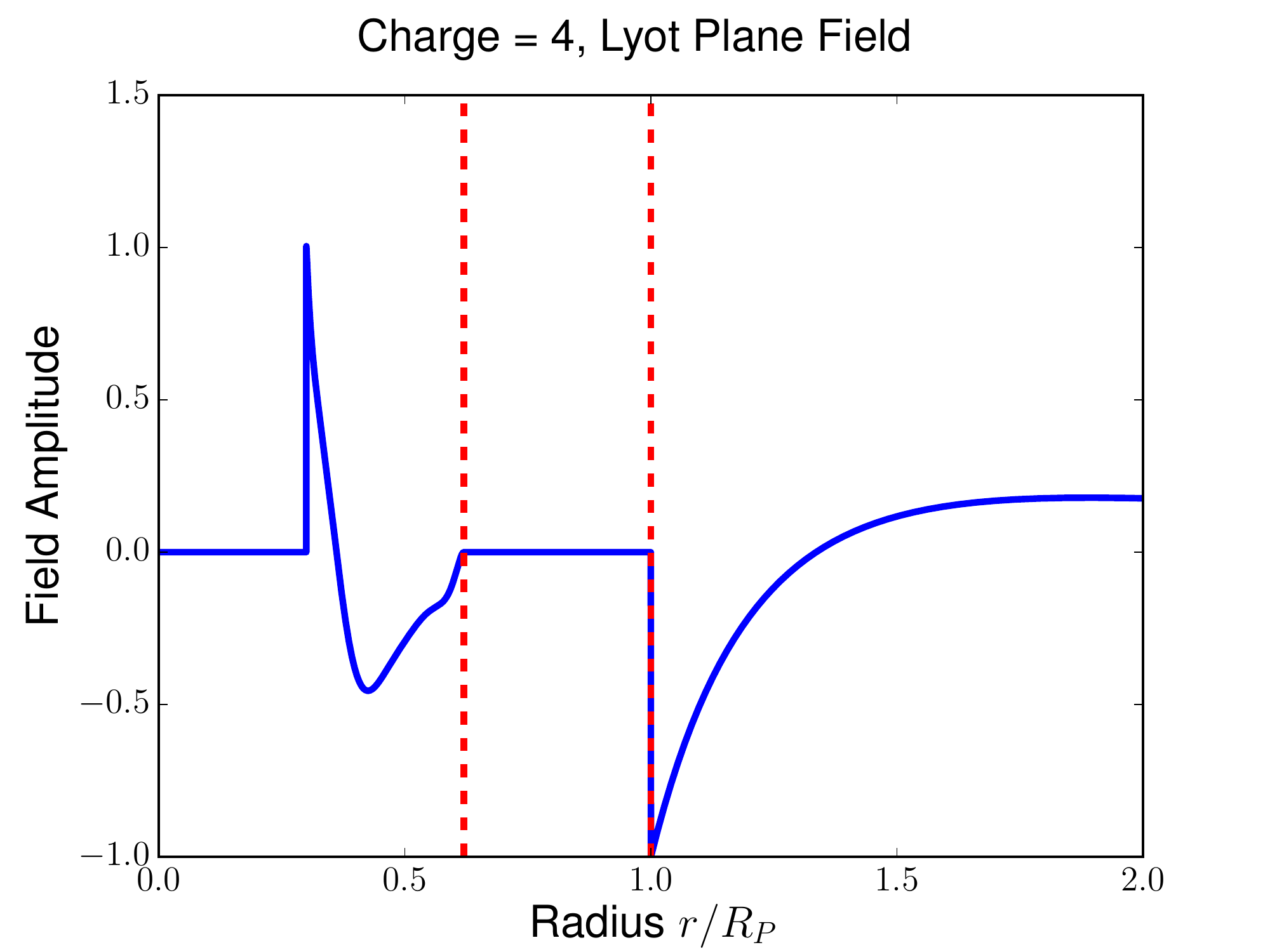}}
~
\subfloat{\includegraphics[height=6.5cm]{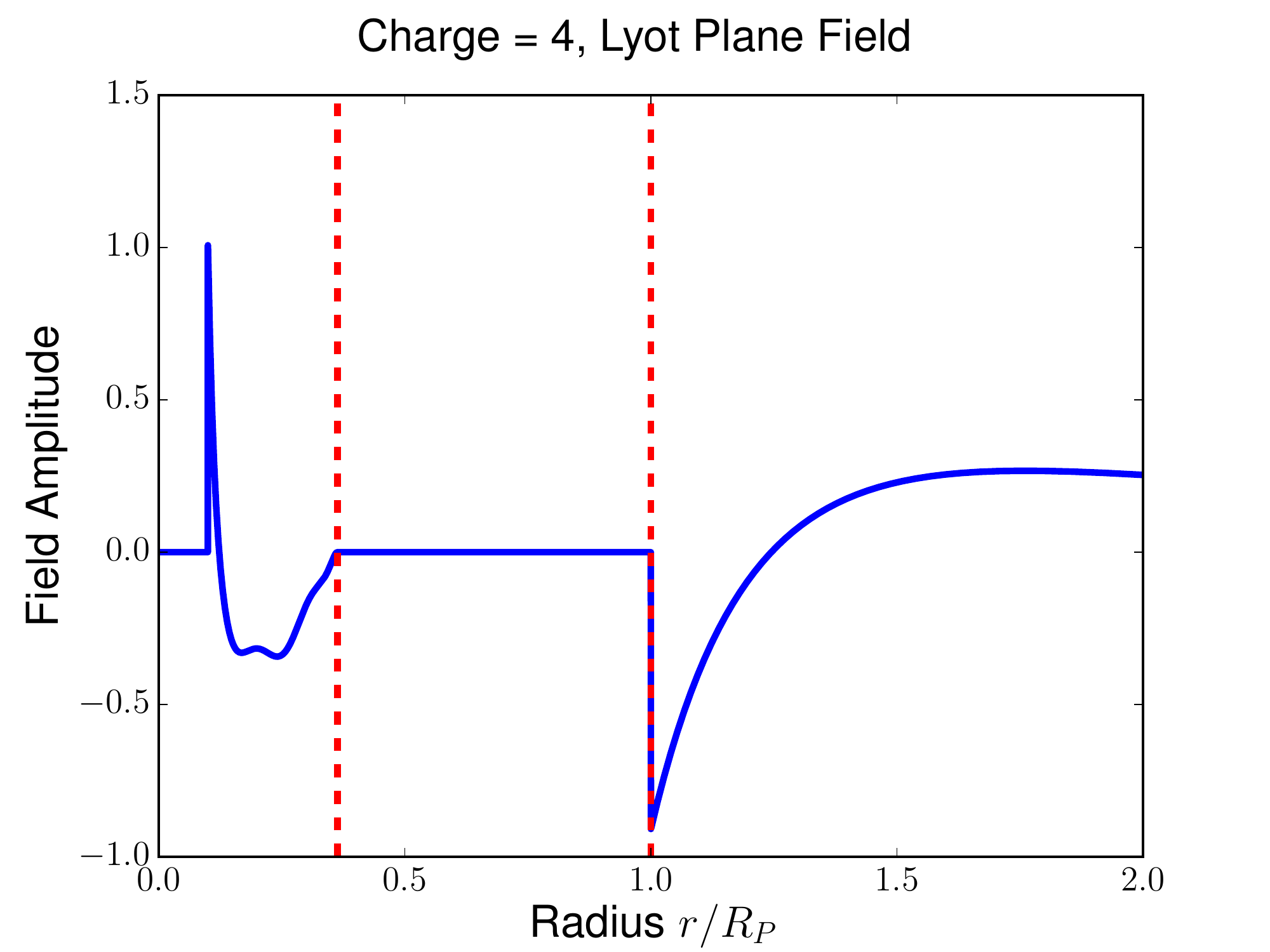}}
\caption[Mirror Shapes]
{ \label{fig:Mirrors} Mirror shape pairs are presented for a charge 4 PAVC for $R_{S} = 0.3R_{P}$ (left) and $R_{S} = 0.1R_{P}$ (right). Mirror shapes are on the top row, and the overall height displacement of the shapes is relative to the sizes and separation of the mirrors. Physical mirror heights depend on the radii of the shaped mirrors $R_{m}$ and the separation $Z$ between the two and go as $R_{m}^{2}/Z$, and translating the relative heights to a physical mirror shape involves multiplying by this factor \citep{2003Traub, 2011Pueyo_Derivs, ACAD}. The apodizations produced by these pairs are shown in the second row, with dashed red lines denoting the radii of the central obscuration and the pupil. The nulled field in the Lyot plane (before the inner Lyot stop) is shown in the third row, with dashed red lines denoting the inner Lyot stop and pupil radii.}
\end{figure*}

\begin{figure*}
\centering
\subfloat{\includegraphics[height=6.5cm]{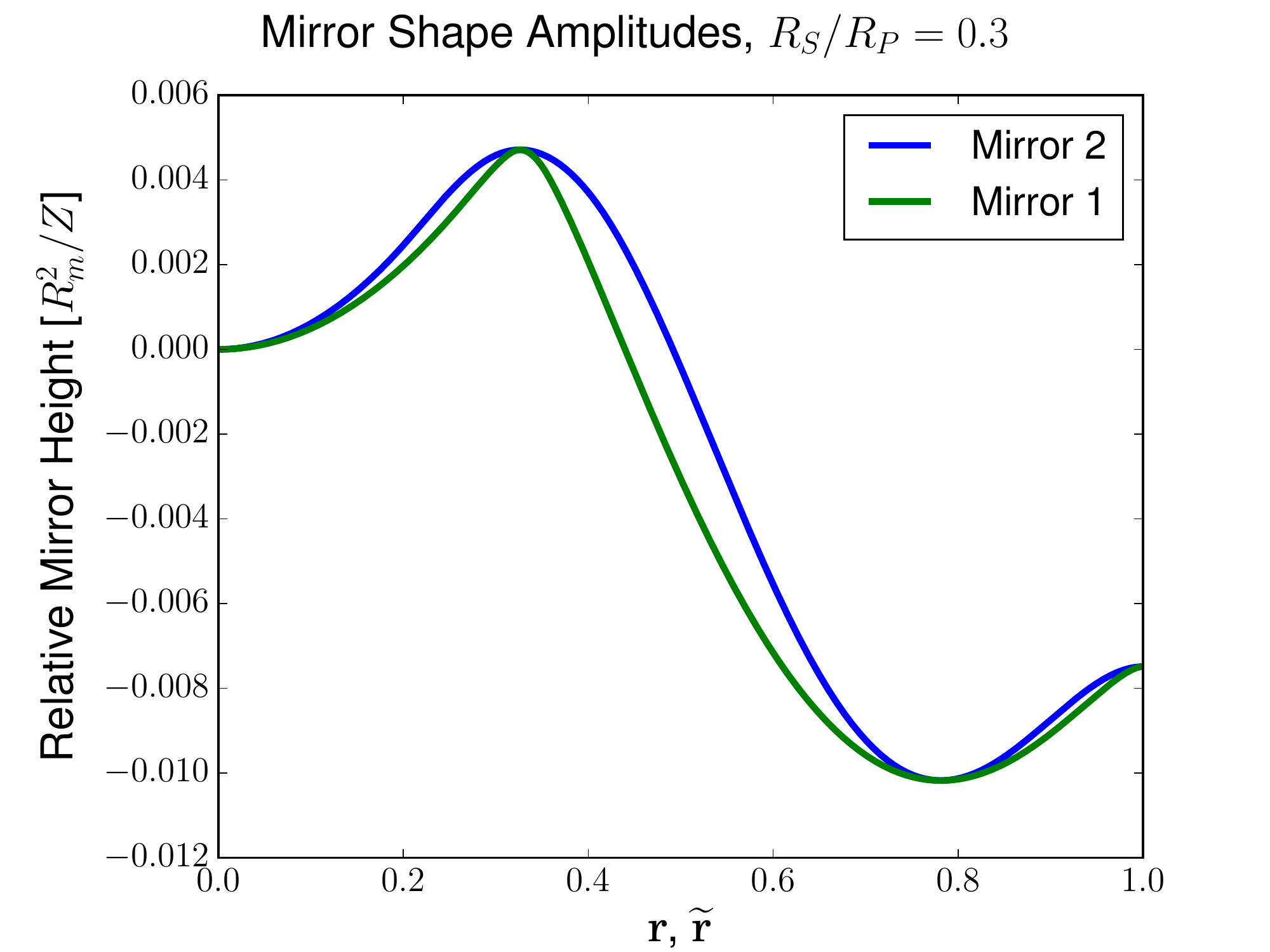}}
~
\subfloat{\includegraphics[height=6.5cm]{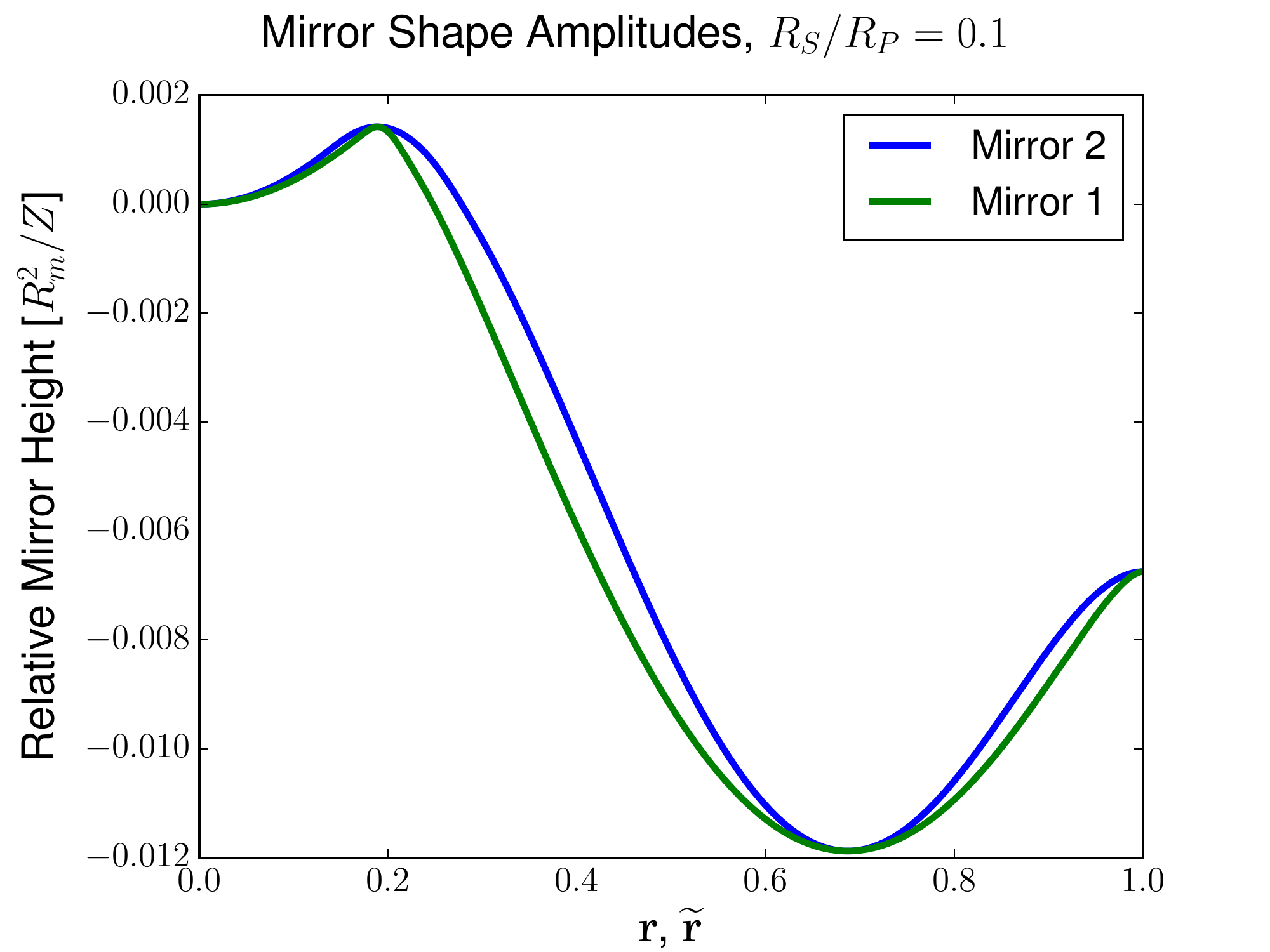}}
\newline
\subfloat{\includegraphics[height=6.5cm]{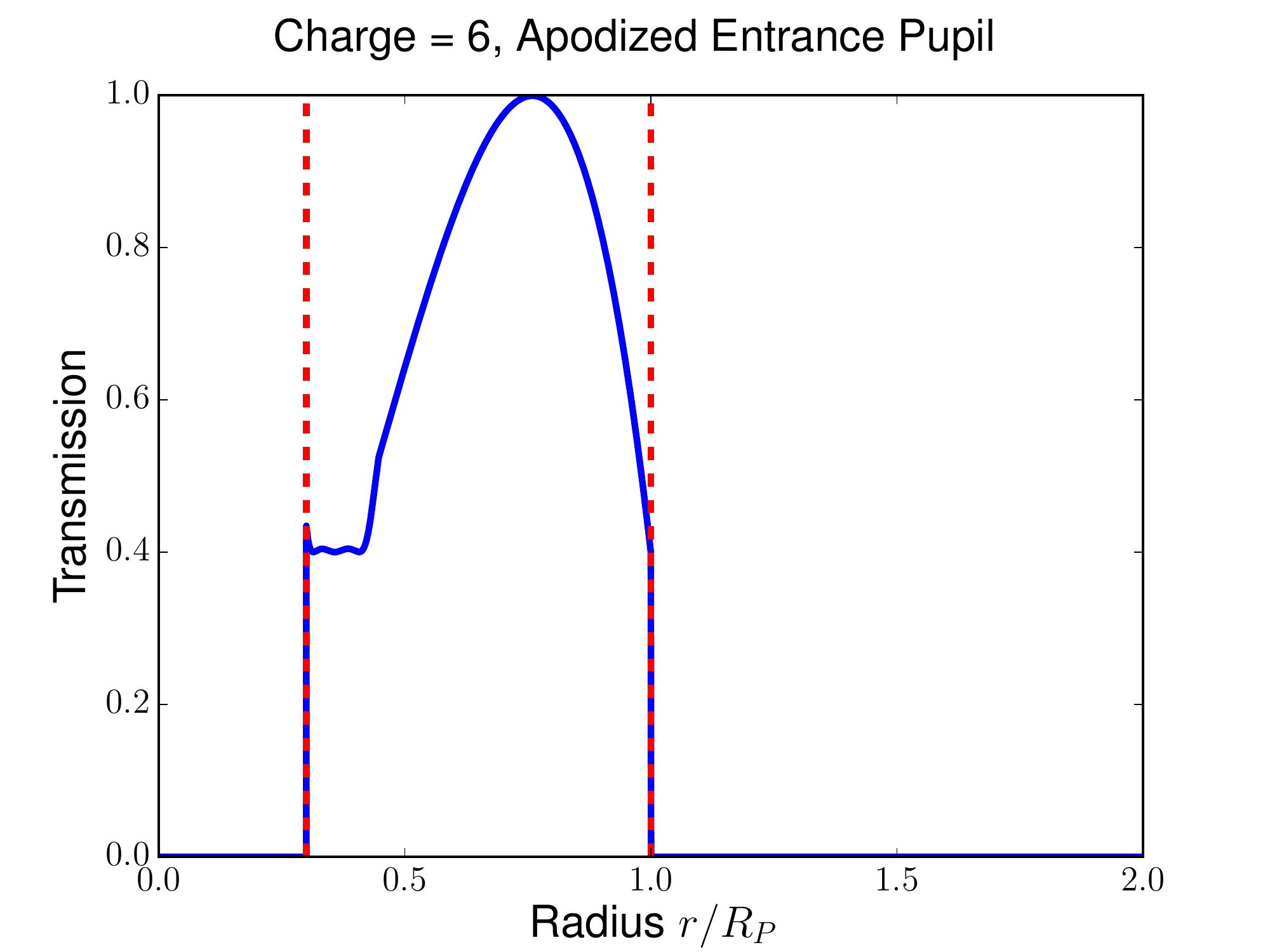}}
~
\subfloat{\includegraphics[height=6.5cm]{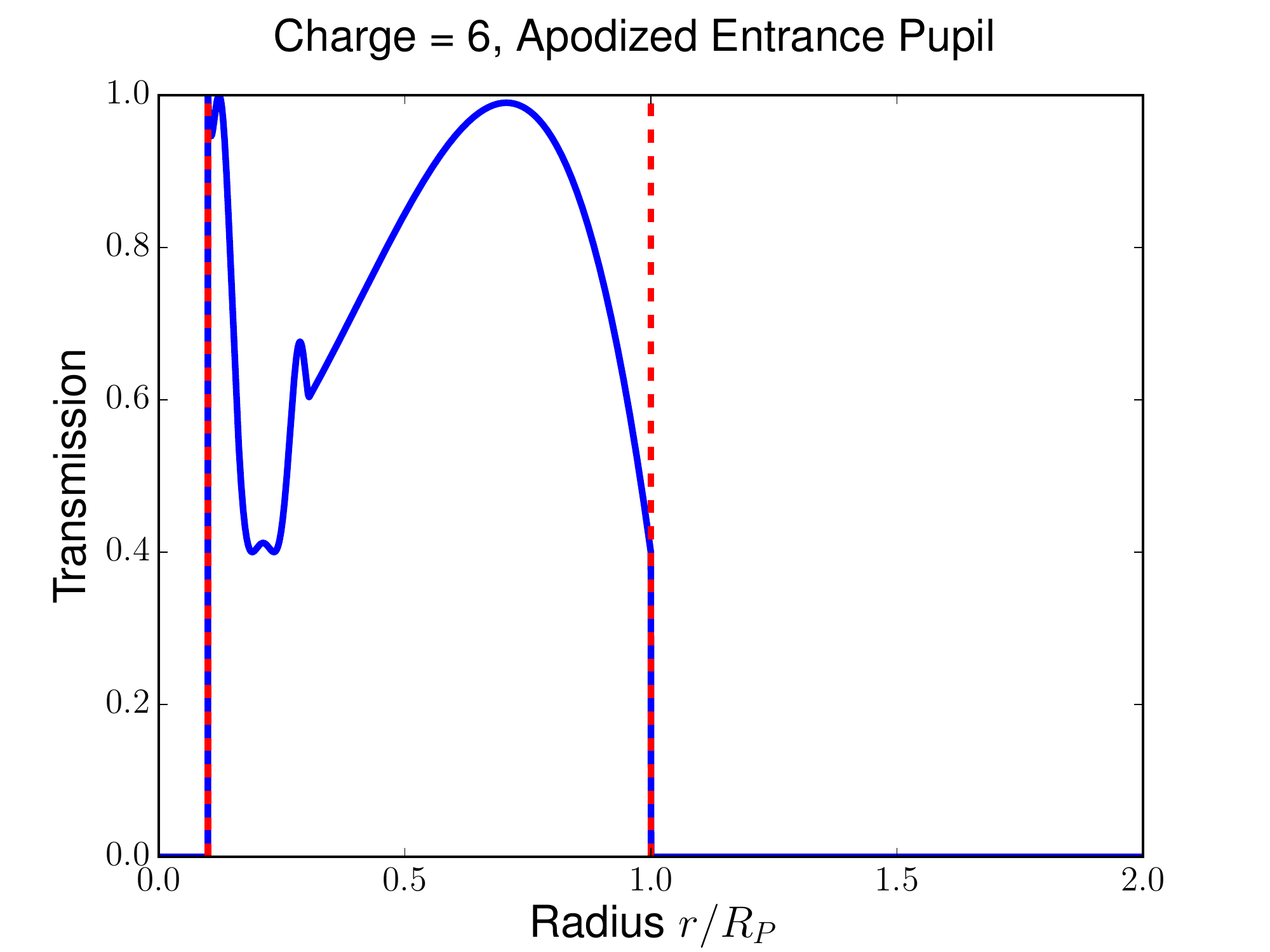}}
\newline
\subfloat{\includegraphics[height=6.5cm]{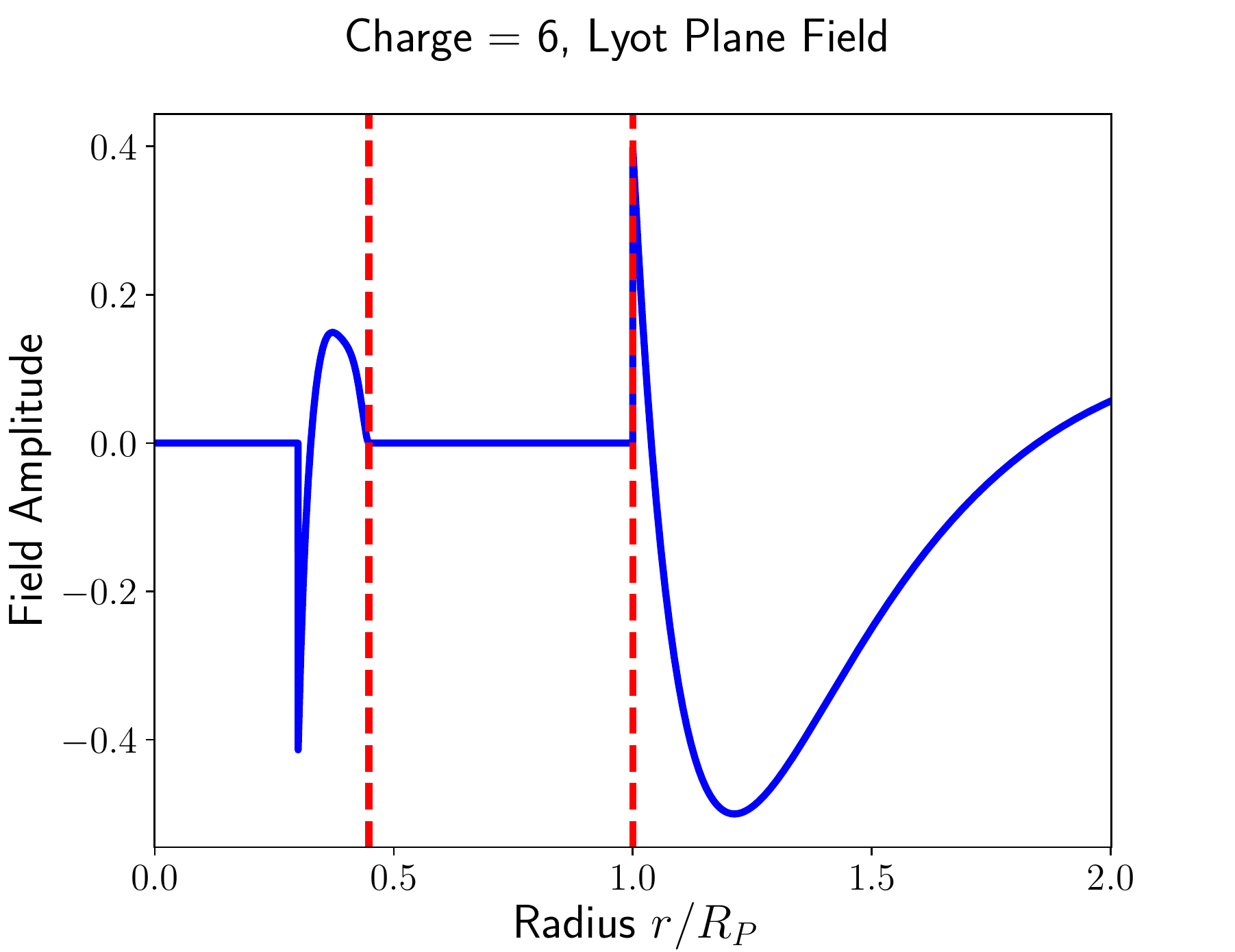}}
~
\subfloat{\includegraphics[height=6.5cm]{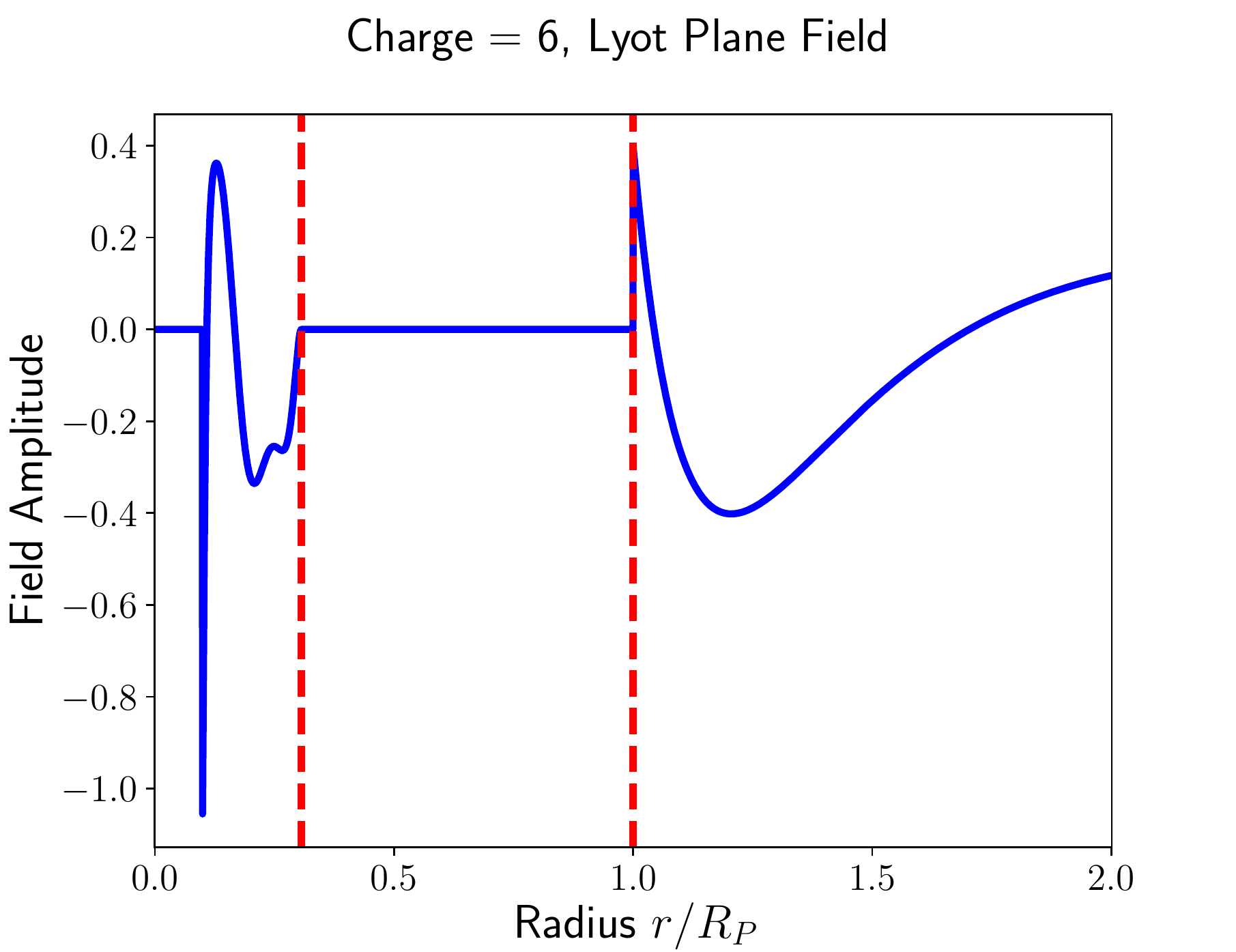}}

\caption[Mirror Shapes]
{ \label{fig:Mirrors_C6} The same as Figure \ref{fig:Mirrors}, except for a charge 6 PAVC for $R_{S} = 0.3R_{P}$ and $R_{S} = 0.1R_{P}$.}
\end{figure*}

Mirror shapes for the charge 4 PAVC are shown in Figure \ref{fig:Mirrors} for $R_{S} = 0.1R_{P}$ and $R_{S} = 0.3R_{P}$. Mirrors for the charge 6 PAVC are shown in Figure \ref{fig:Mirrors_C6}. Since mirror shapes are normalized to conserve energy, the total energy throughput is:
\begin{equation}\label{eq:26}
T.E. = \frac{\pi\left(R_{P}^{2}-R_{I}^{2}\right)}{\pi\left(R_{P}^{2}-R_{S}^{2}\right)}.
\end{equation}
For the charge 4 PAVCs in Figure \ref{fig:Mirrors}, the total energy throughput is $88\%$ for $R_{s} = 0.1R_{P}$ and $68\%$ for $R_{S} = 0.3R_{P}$. For the charge 6 PAVCs total energy throughput is $86\%$ and $88\%$, respectively.

\begin{figure*}
{\centering
\subfloat{\includegraphics[height=6.125cm]{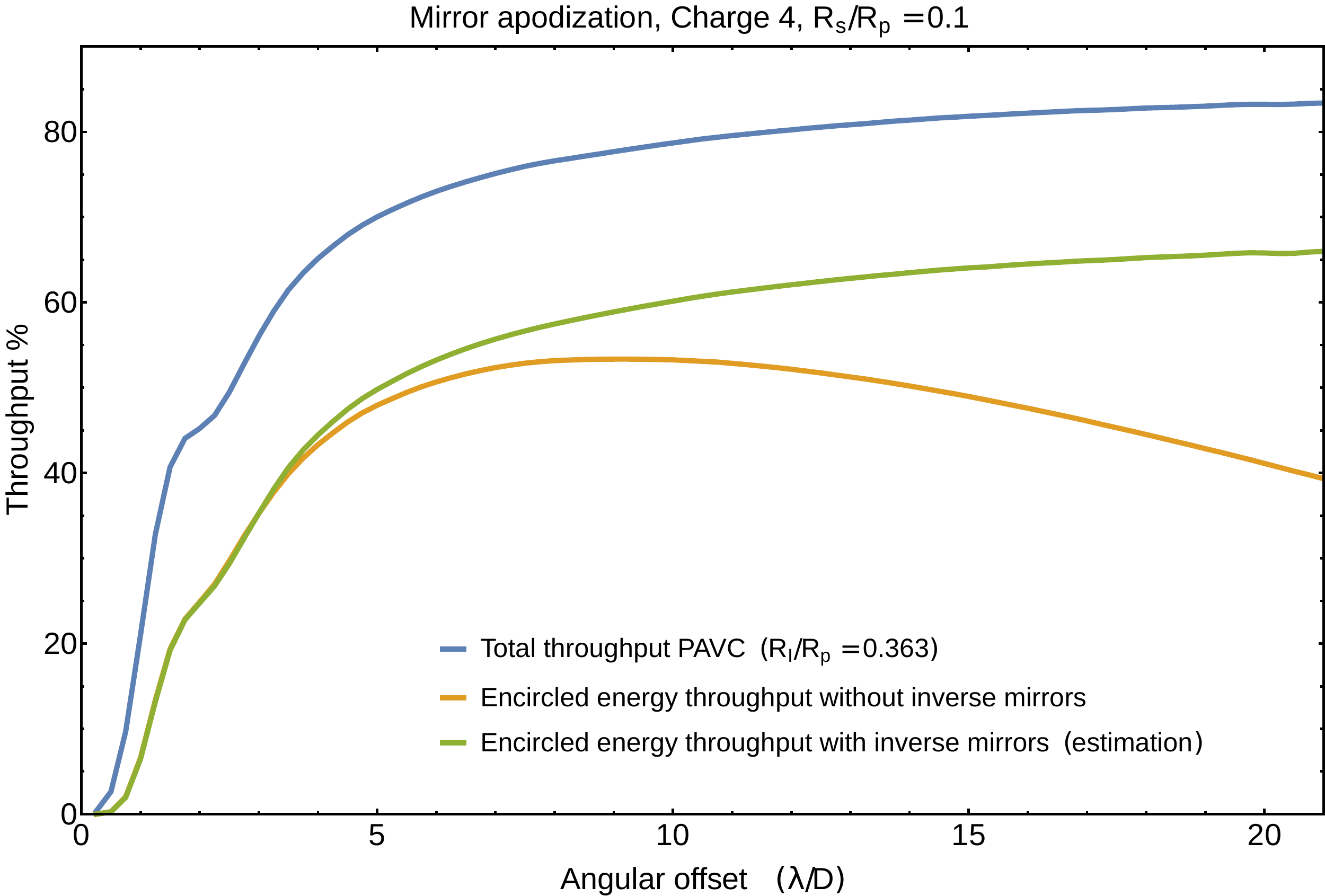}}
~
\subfloat{\includegraphics[height=6.125cm]{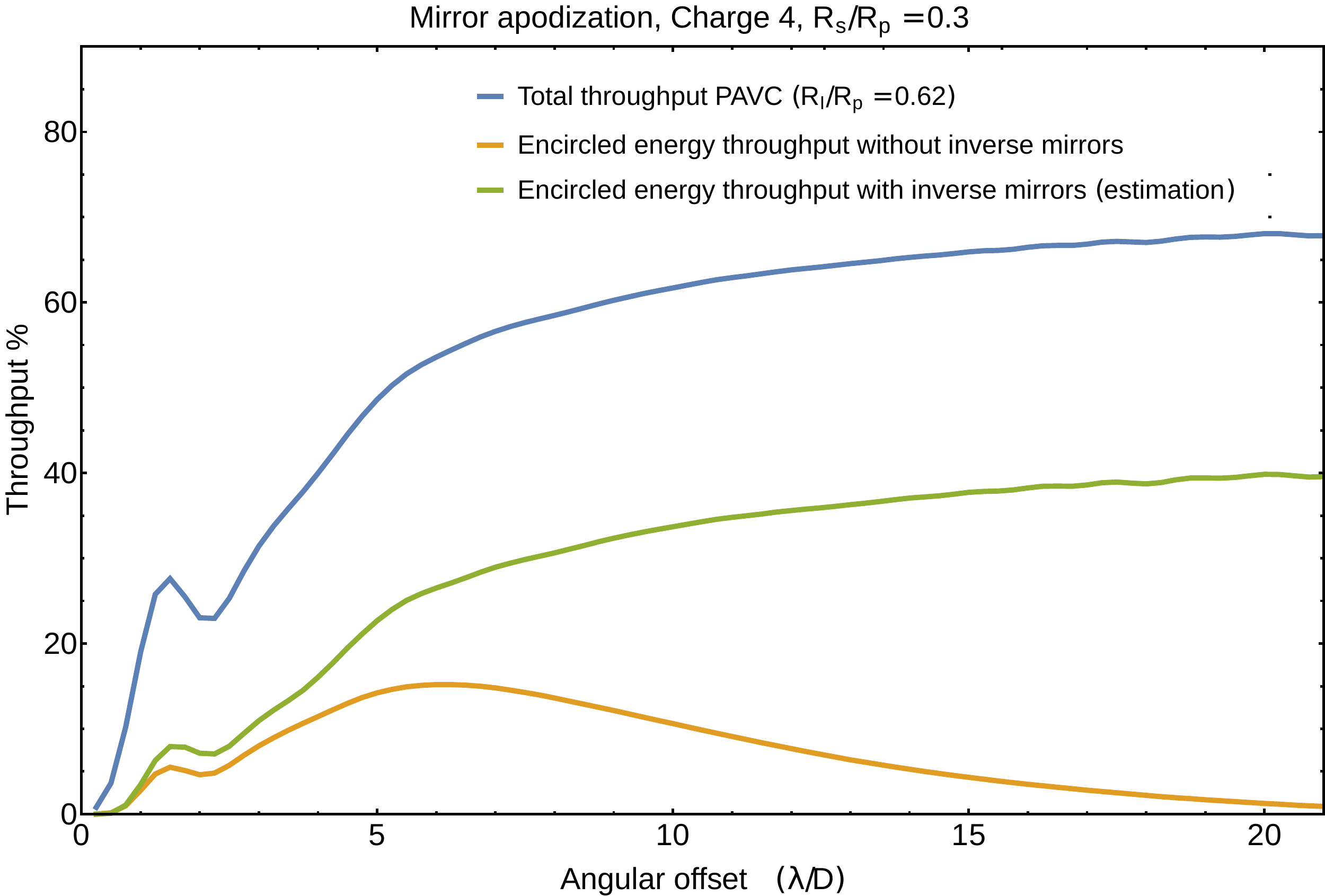}}
\newline
\subfloat{\includegraphics[height=6.125cm]{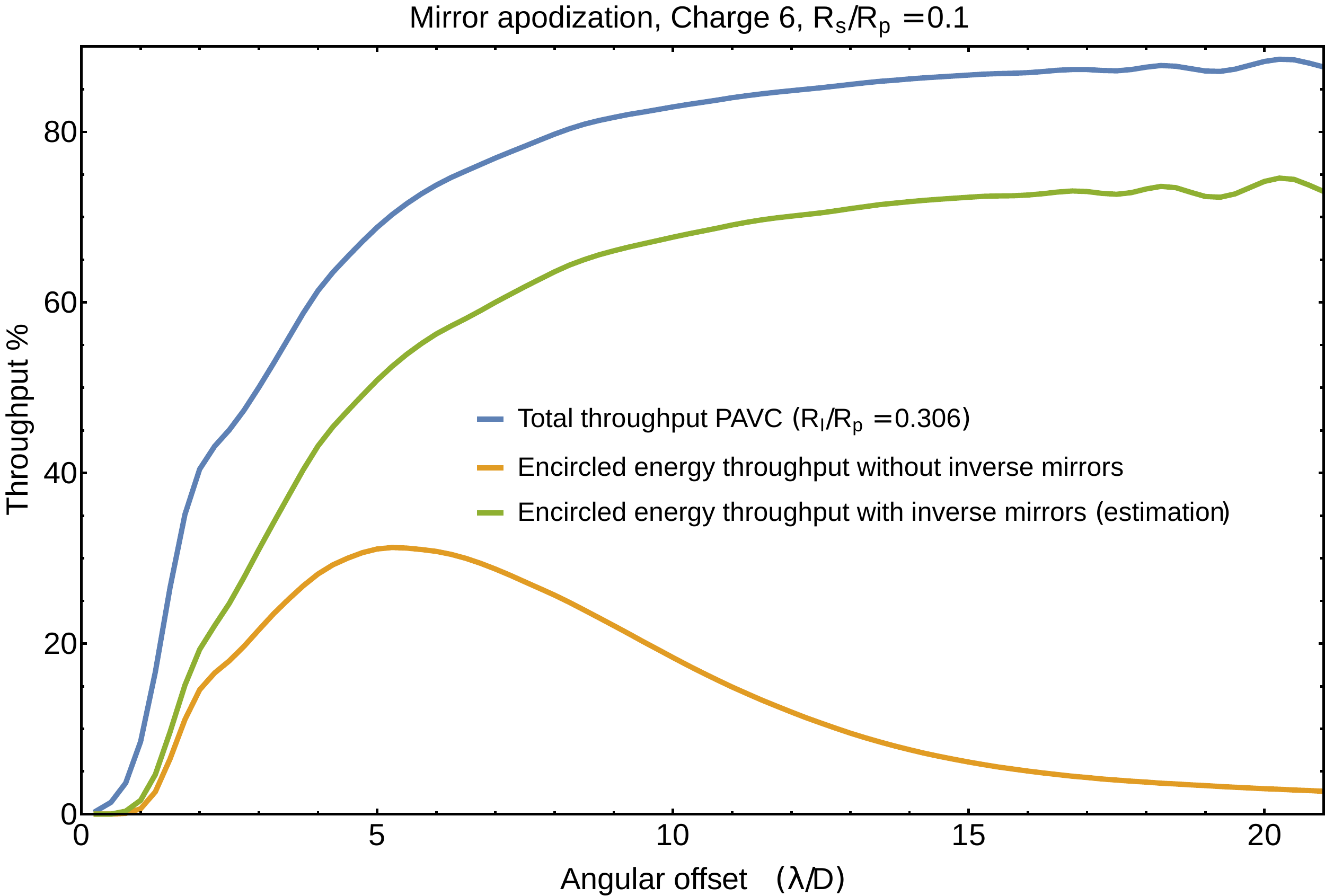}}
~
\subfloat{\includegraphics[height=6.125cm]{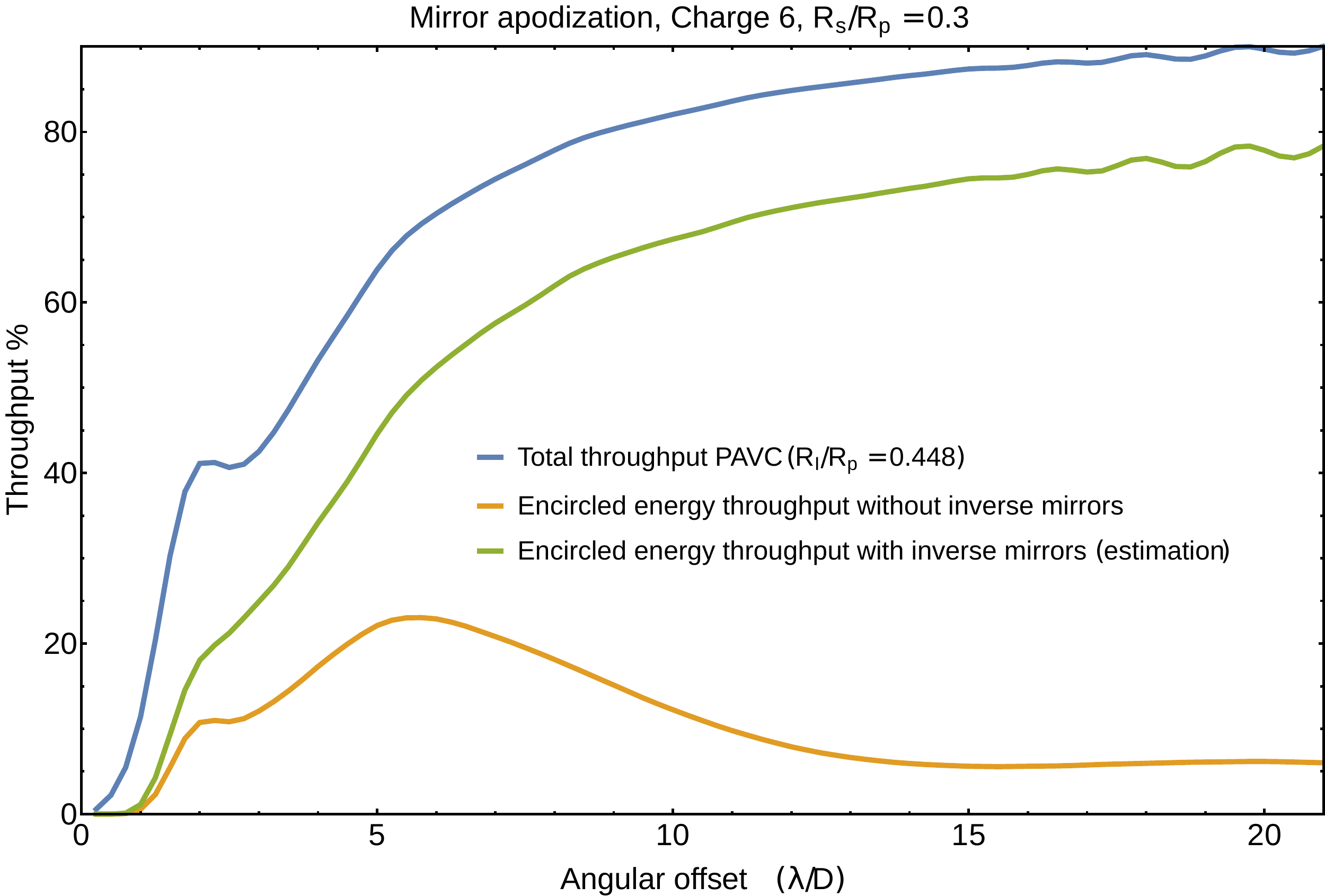}}}

\caption[Mirror Shapes]
{ \label{fig:Off_Axis_Mirrors} \textit{Top Row:} Off-axis throughputs for optimal coronagraph designs with a charge 4 PAVC, where apodizations were obtained with shaped mirrors. The left-hand panel shows the throughput of an off-axis source, as a function of distance from the on-axis star in units of $\lambda$/D for a coronagraph designed for a telescope with $R_{S} = 0.1R_{P}$. The blue line is the total throughput, and the yellow line is the encircled energy throughput. The green line depicts the approximate encircled energy throughput obtained when the off-axis PSF is restored by inverse mirrors placed after the Lyot plane. The right-hand panel show the same set of curves, this time for a coronagraph designed for a telescope with $R_{S} = 0.3R_{P}$.  \textit{Bottom Row:} The same set of plots as the top two rows, but for a coronagraph designed with a charge 6 vortex.}
\end{figure*}

The off-axis throughput curves for shaped-mirror PAVCs are shown in Figure \ref{fig:Off_Axis_Mirrors}. Total energy throughputs are shown as blue curves, and demonstrate a gain in throughput over the use of a greyscale apodizer for each combination of charge and secondary mirror radius we examined. Encircled energy throughputs are denoted by yellow curves. For each shaped-mirror PAVC, except for the charge 4 PAVC with secondary mirror radius $R_{S}/R_{P} = 0.1$, the mirror shapes we calculate severely distort the off-axis point spread function (PSF) and result in a narrow range of angular separations from the on-axis star where even a modest encircled energy throughput can be obtained. Therefore, we propose using inverse mirrors placed after the Lyot stop in Stage C in Figure \ref{fig:Mirror_Overview}, similar to the inverse optics proposed for PIAA mirrors \citep{2005Guyon_PIAA, 2014Guyon}. Our estimate of the encircled energy throughput after the off-axis PSF is restored by inverse optics is shown by the green curve for each coronagraph design in Figure \ref{fig:Off_Axis_Mirrors}. 

We estimated restored off-axis PSFs by propagating the off-axis point source flux through the coronagraphic setup without a pair of shaped mirrors to reshape the pupil, and with the total energy normalized by the total energy in the final image plane of the apodized coronagraph. The restored encircled energy throughputs we present are therefore upper limits on what may be achieved with inverse optics. If we treat the encircled energy throughput of an off-axis source as the best estimate for the `usable' photometric flux, then using shaped mirrors with inverse optics represents an improvement over greyscale filters for the charge 4 PAVC by a factor of 2 and the charge 6 PAVC by about 50$\%$.

\section{Application To Future Telescopes}

Moving forward, the PAVC formalism can be used in the design of instruments for arbitrary telescope pupils. In combination with techniques for minimizing the impact of pupil features imposed by secondary mirror support struts and primary mirror segmentation gaps, the PAVC enables vortex coronagraphs to be used with the complicated pupils of future space and ground based large-aperture on-axis telescopes. Eventually, it will be necessary to extend the PAVC formalism to address the sensitivity of obstructed vortex coronagraphs to low-order aberrations, and to define and optimize figures of merit that are best suited to real-world on-axis telescopes. However, by effectively solving the problem posed by central obscurations, the PAVC provides the basis for designing vortex coronagraphs for on-axis telescopes.

 Recently, numerical algorithms have been put forward to calculate either apodizing mask or shaped mirror geometries that mitigate the impact of struts and primary mirror gaps (so-called `spiders') on vortex coronagraph starlight suppression \citep{2016Ruane_Apod, 2015Mazoyer_ACAD, 2016Mazoyer_OSMSPIE}. By using a PAVC apodizer to correct the central obscuration, we are able to provide a first order correction for the vortex coronagraph on a pupil obstructed by both a secondary mirror and spiders. Higher order corrections to the PSF can then be obtained using either e.g. \cite{2016Ruane_Apod} or \cite{2016Mazoyer_OSMSPIE} to create a dark hole with $\sim 10^{-10}$ contrast while minimizing losses to throughput. 

\begin{figure*}
\begin{center}
\begin{tabular}{c}
\includegraphics[height=12cm]{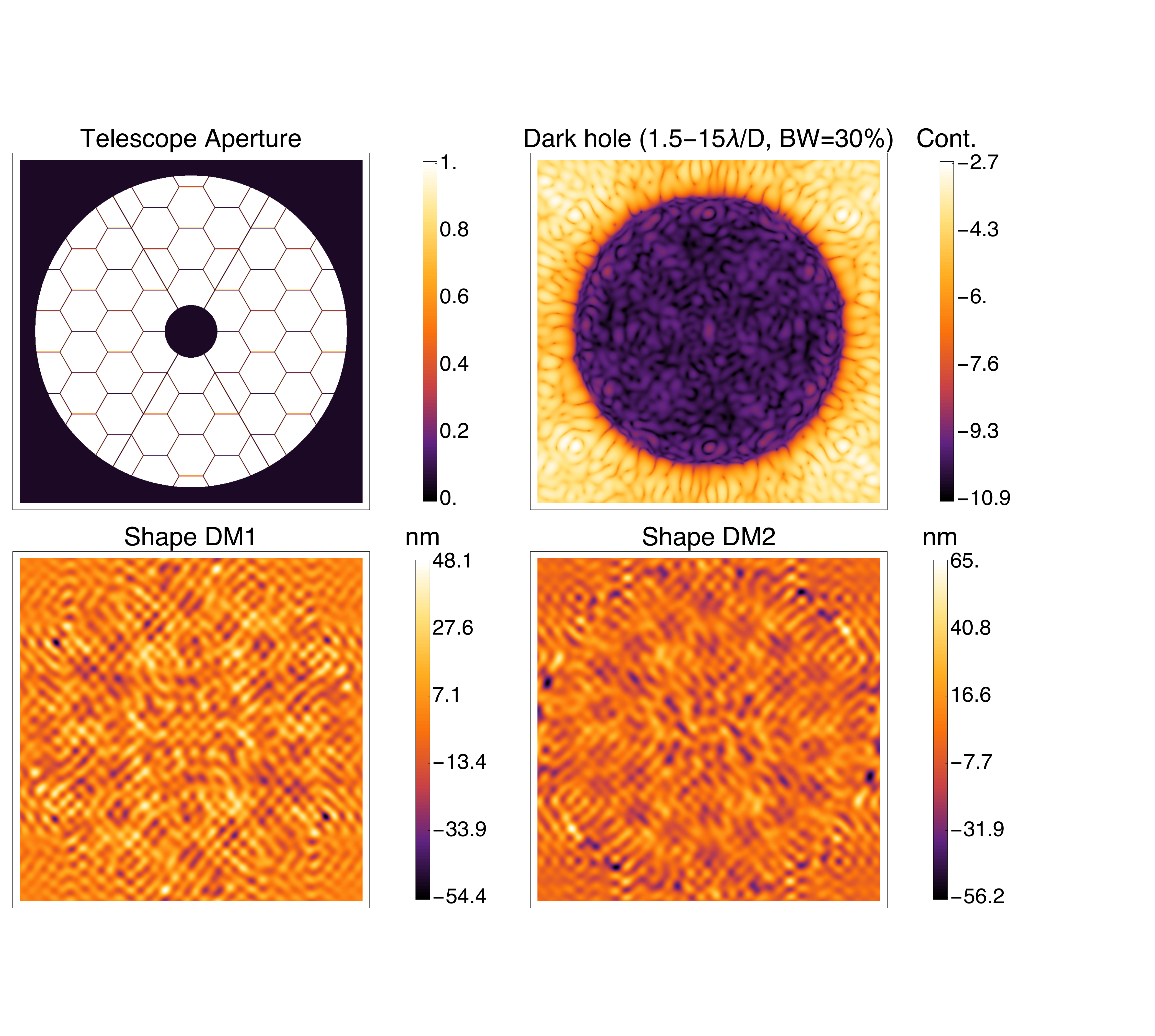}
\end{tabular}
\end{center}
\caption[]
{ \label{fig:Obs_Pupil} An example deformable mirror (DM) solution that creates a dark hole PSF for a centrally obstructed aperture in conjunction with a polynomial apodizing mask. The pupil, featuring a secondary mirror obscuration with $R_{S}/R_{P} = 0.17$ and segment gaps is shown in the upper left. The lower left and lower right show the mirror shapes that provide the higher-order correction for segment gaps using ACAD-OSM \citep{2016Mazoyer_OSMSPIE} and, in conjunction with an apodizing mask, produce the PSF in the upper right. The PSF features a dark hole with a contrast of approximately $10^{-10}$ in the region $1.5-15$ $\lambda$/D. }
\end{figure*}

Figure \ref{fig:Obs_Pupil} depicts an example of how a such a combined approach works on a pupil with a central obscuration of radius $R_{S}/R_{P} = 0.17$ and spiders. The pupil shown in Figure \ref{fig:Obs_Pupil} is a realistic aperture for future space telescope missions taken from the Segmented Coronagraph Design and Analysis (SCDA) program.\footnote{https://exoplanets.nasa.gov/system/internal\_resources/details/ \\ original/211\_SCDAApertureDocument050416.pdf} Using a charge 6 PAVC with an inner Lyot stop of radius $R_{I} = 0.414R_{P}$, we correct for the central obscuration. In the absence of spiders, the PAVC alone provides an encircled energy throughput of $44\%$. With the addition of spiders, the PAVC alone has a contrast ratio of only $10^{-6}$ assuming broadband light with a $30\%$ bandwidth. 

When we combine the apodizing filter for the PAVC with mirror shapes obtained using the ACAD-OSM (Active Correction of Aperture Discontinuities- Optimized Stroke Minimization) algorithm described in \cite{2016Mazoyer_OSMSPIE}, we obtain the PSF shown in the upper right portion of Figure \ref{fig:Obs_Pupil}. ACAD-OSM builds on the deformable mirror-based Active Correction of Aperture Discontinuities (ACAD) algorithm presented in \cite{ACAD}. 

\begin{figure}
\begin{center}
\begin{tabular}{c}
\includegraphics[height=6cm]{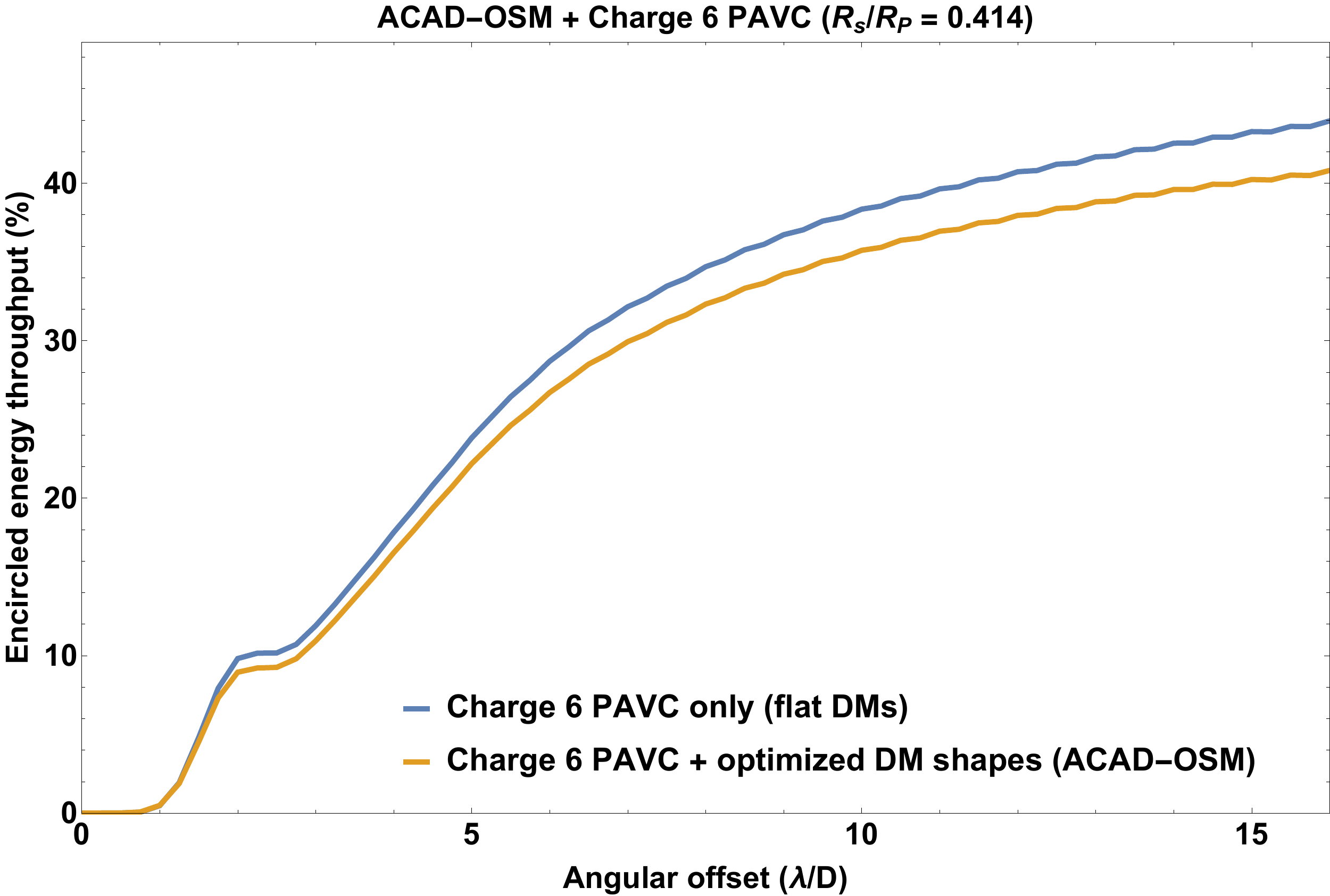}
\end{tabular}
\end{center}
\caption[]
{ \label{fig:Spider_Throughput} Encircled energy throughput vs. offset for a polynomial apodized + ACAD-OSM vortex coronagraph design. The blue curve shows the throughput obtained in the absence of deformations on the deformable mirrors (DMs), while the purple curve shows the throughput obtained when both a polynomial apodizer and shaped mirrors are used to correct the central obscuration and segment gaps.}
\end{figure}

The shaped mirror corrections provide a PSF with a dark hole between 1.5 and 15 $\lambda$/D with a contrast of $10^{-10}$ in $30\%$ bandwidth light. Encircled energy throughput of an off-axis point source as a function of angular separation from the star is shown in Figure \ref{fig:Spider_Throughput}. The additional correction to the pupil geometry provided by the shaped mirrors reduces encircled energy throughput by only a few percent, and the encircled energy throughput in the dark hole exceeds $20\%$ at 4.5 $\lambda$/D. At the outer edge of the dark hole, where throughput is highest, this rises to $\sim 41\%$. 

The example coronagraph design discussed here provides proof of concept for combining the PAVC with algorithms that compensate for the effects of struts and segments on telescope pupils. It may be optimized substantially depending on demands imposed on the coronagraph by the initial pupil geometry. In particular, throughput of the spider correction depends heavily on the inner Lyot stop radius used, and it may be possible to determine an ideal inner Lyot stop radius for a combination of polynomial apodizing filter and spider-correcting apodization. Furthermore, polynomial apodizing geometries that correct for the central obscuration may also be optimized to result in a pupil geometry that requires less correction with numerical techniques.

Application of the PAVC to real-world obstructed pupils with spiders will likely require us to optimize the throughput of a combined PAVC + spider correcting apodization design. As discussed in $\S$ 4, we calculated PAVC apodizations by maximizing transmission through the apodizing filter, which is a linear quantity that approximates throughput. However, the actual quantity that determines the relative flux from an exoplanet or other off-axis source that the observer will be able to detect is best approximated by the encircled energy throughput shown in Figures \ref{fig:Off_Axis}, \ref{fig:Off_Axis_Mirrors}, and \ref{fig:Spider_Throughput}. Since the encircled energy throughput is a function of angular separation, the optimal FOM will depend on performance goals for effective inner working angle as well as overall throughput. For example, the optimal FOM for a given coronagraph may be the encircled energy throughput at 3 $\lambda$/D.  Such a FOM will require us to adopt a non-linear algorithm for optimizing the pupil apodization. Fortunately, since PAVC designs have a large number of degrees of freedom compared to the contraints needed to ensure ideal starlight suppression, they are well-suited to addressing complex optimization problems.

Our derivation of analytical expressions for $V_{c}\left[A\left(r\right)\right]$ also raises the possibility of creating PAVC designs that are highly robust to low-order aberrations. Currently, we are working on extending our technique for deriving PAVC apodizations to analytically describe impacts of low-order aberrations on PAVC designs, and to optimize the PAVC to minimize sensitivity to these effects. With an unobstructed circular aperture, the sensitivity of a vortex coronagraph to small angular offsets of the on-axis source caused by low-order aberrations goes approximately as $s^{c}$, where $s$ is angular separation of the offset between the telescope pointing and aberrant flux from the on-axis source (typically $\ll 1$ $\lambda$/D) and $c$ is the charge of the vortex \citep{2008Jenkins, 2010Mawet_Aberrations}. Therefore, as the charge of a vortex coronagraphs increases, the coronagraph becomes more robust to low-order aberrations. However, in the presence of a central obscuration, vortex coronagraphs are less robust to these aberrations for a given charge, and the flux `leak' caused by small angular offsets rises more steeply than $s^{c}$. Since the PAVC allows for high-throughput coronagraphs be constructed for on-axis telescopes with charge 4 and 6 vortices, addressing this issue of robustness will allow the PAVC to take advantage of the stability of these charges relative to the charge 2 vortex, as well as their relative insensitivity to finite stellar angular size.

\section{Conclusion}

The analytical prescription we derived for the PAVC overcomes a fundamental hurdle associated with the vortex coronagraph-- the loss of starlight suppression due to the obscuration created by the secondary mirrors of on-axis telescopes. With the apodizing pupil masks in this paper, we are able to restore the theoretically total and achromatic starlight suppression of the vortex coronagraph while maintaining off-axis throughput. When produced with a classically apodizing mask, the apodization functions we derive offer throughputs that are competitive with or substantially higher than other high-contrast coronagraph designs \citep[e.g.][]{2013Mawet_RAVC, 2015NDiaye_APLC, 2014Guyon} for centrally obscured pupils, especially with a charge 6 vortex. When produced with shaped mirrors, PAVC throughput can be improved even further, albeit at the cost of a more complicated instrument design.

The classically apodized PAVC has a total energy throughput of $\sim$75\%, 75\%, and 80\% for charges 2-6 respectively with a central obscuration with $R_{S} = 0.1R_{P}$, and $\sim$ 35\%, 45\%, and 70\% with a central obscuration with $R_{S} = 0.3R_{P}$. Predictably, total energy throughput decreases as the radius of the central obscuration increases. However, the apodizations produced with our PAVC formalism perform better for all definitions of throughput discussed in this paper as the vector vortex charge increases, particularly with large central obscurations. For a central obscuration of $R_{S} = 0.3R_{P}$, the encircled energy throughput is $\sim 22\%$ for the charge 4 PAVC, dropping from $\sim 50\%$ when $R_{S} = 0.1R_{P}$. For the charge 6 PAVC, total energy throughput is $\sim 50\%$ when $R_{S} = 0.3R_{P}$, and only slightly higher for smaller central obscurations.

Using shaped mirror pairs to apodize the PAVC allows us to further improve throughput, but requires a more complicated design. We found mirror shapes for charge 4 and 6 PAVCs. For the case of the charge 4 PAVC, we obtained a total energy throughput of $88\%$ with $R_{S} = 0.1R_{P}$ and $68\%$ with $R_{S} = 0.3R_{P}$. For the charge 6 PAVC, these figures are $86\%$ and $88\%$, respectively. Since the shaped mirrors induce considerable distortion to the off-axis PSF, in order to obtained reasonable encircled energy throughputs for off-axis sources, it is necessary to add inverse optics to the PAVC downstream of the Lyot plane. With inverse optics, designing the PAVC with shaped mirrors instead of an apodizing filter could improve the encircled energy throughput of the charge 4 PAVC by as much as a factor of 2, and the charge 6 PAVC by 50\%. While the PAVCs we present use mirror shapes that require inverse-mirror optics to obtain useful off-axis PSFs, our algorithm can also be used to find the trade-off between throughput and curvature that may allow us to overcome the limitation imposed by off-axis PSF distortion by the shaped mirrors.

Our results overcome the challenge posed to vortex coronagraphs by large central obscurations. We describe a formalism for finding pupil apodizations that restore ideal starlight suppression while preserving high throughputs for off-axis sources for vortex coronagraphs with charges $>$ 2. As mentioned in \cite{2016Ruane_Apod}, synergy exists between apodized vortex coronagraph designs that compensate for central obscurations and designs that compensate for pupil obstructions introduced by struts and primary mirror gaps. We provide an example of implementing a synergistic PAVC + numerical design by combining the PAVC with ACAD-OSM pupil shaping \citep{2016Mazoyer_OSMSPIE}. Apodized vortex coronagraphs such as this example are promising candidates for high-contrast instruments designed for on-axis telescopes with complicated pupils. Our results greatly expand the flexibility of vortex coronagraph designs, and represent a step forward towards implementing vortex coronagraphs capable of directly imaging Earth-sized exoplanets.

\section*{Acknowledgments}

We acknowledge discussions with Garreth Ruane and Dimitri Mawet that helped enhance our discussion on combining the PAVC with numerical apodization techniques to design coronagraphs for telescopes with pupils that feature both central obscurations and `spiders.' This material is partially based upon work carried out under subcontract 1496556 with the Jet Propulsion Laboratory funded by NASA and administered by the California Institute of Technology.


\bibliographystyle{apj}
\bibliography{VVC_Refs}

\appendix
\section{Vortex Operator for Analytical Pupil Shapes}

While the central obscuration imposed by the secondary mirror is the largest feature in the pupil geometry of an on-axis telescope design, on axis designs will also feature a number of spiders. In the following, we demonstrate how the prescription we present in Section 2 for circularly symmetric pupils may be extended to describe the propagation of non-circularly symmetric components of a telescope pupil.

The vortex operator we have described in this paper may be extended for any aperture geometry that can be described by a Zernike polynomial expansion. Such a description allows us to analytically propogate the effects of both the secondary mirror of an on-axis telescope and the pupil `spiders' which result from discontinuities in the primary mirror and the support structure for the secondary mirror. 

We consider a pupil geometry which can be expanded into terms $r^{n_{p}}e^{im_{p}\theta}$, in the region $R_{S}\leq r\leq R_{P}$, and an apodizing mask which can be expanded into terms $r^{n_{m}}e^{im_{m}\theta}$. The apodized pupil geometry therefore consists of terms 
\begin{equation}
r^{n}e^{im\theta} = \left(r^{n_{p}}e^{im_{p}\theta}\right)\left(r^{n_{m}}e^{im_{m}\theta}\right).
\end{equation}
 The electric field at the Lyot plane for a $c=4$ coronagraph for one of these terms is analytic so long as $n + m$ is even, unless $n \geq 0$ and $n + m = -4$ or $n + m = -2$. For a given term
\begin{equation}\label{eq:PupTerm}
f_{n,m}\left(r,\theta\right) = \begin{cases}
  0, & \text{if $r<R_{i}$}. \\
  r^{n}e^{im\theta}, & \text{if $R_{i}\leq r\leq R_{P}$} \\
  \end{cases}
\end{equation}
the electric field in the Lyot plane at $R_{i}\leq r\leq R_{P}$ is
\begin{equation}\label{eq:LyotTerm}
V_{4}\left[f_{n,m,p}\left(r,\theta\right)\right] = 
  e^{i\left(m+4\right)}\left[C_{0}\left(n,m\right)r^{n}+ C_{1}\left(n,m\right)r^{-m-2}+ C_{2}\left(n,m\right)r^{-m-4}\right]
\end{equation}
where
\begin{equation}\label{eq:C0nm}
C_{0}\left(n,m\right) = \frac{\left(n-m\right)\left(n-m-2\right)}{\left(n+m+4\right)\left(n+m+2\right)},
\end{equation}
\begin{equation}\label{eq:C1nm}
C_{1}\left(n,m\right) =  \frac{-2\left(m+1\right)\left(m+2\right)}{m+n+2} \begin{cases}
  R_{i}^{m+n+2}, & \text{if $m \geq -1$} \\
  R_{P}^{m+n+2}, & \text{if $m < -1$}, \\
  \end{cases}
\end{equation}
and
\begin{equation}\label{eq:C2nm}
C_{2}\left(n,m\right) = \frac{2\left(m+2\right)\left(m+3\right)}{m+n+4} \begin{cases}
  R_{i}^{m+n+4}, & \text{if $m \geq -1$} \\
  R_{P}^{m+n+4}, & \text{if $m < -1$}. \\
  \end{cases}
\end{equation}

\end{document}